\documentclass[12pt]{article}
\usepackage{graphicx}
\usepackage{amsmath}
\usepackage{amsfonts}
\usepackage{amsthm}
\usepackage{enumerate,setspace}
\usepackage{bbm,comment}
\usepackage{authblk}
\usepackage{dcolumn,booktabs}
\usepackage{siunitx}

\usepackage[natbibapa]{apacite}
\usepackage{listings}

\usepackage{xcolor}
\usepackage{hyperref}
\usepackage{enumitem}
\usepackage{minted}
\usepackage{xcolor}
\definecolor{LightGray}{gray}{0.9}
\usepackage{multirow}
\usepackage{placeins}
\usepackage{float}

\newcommand{\blind}{0}

\definecolor{comcolor}{rgb}{0,0.5,0}
\definecolor{funcolor}{rgb}{0,0.4,1}
\definecolor{concolor}{rgb}{0,0,1}

\definecolor{codegreen}{rgb}{0,0.6,0}
\definecolor{codegray}{rgb}{0.5,0.5,0.5}
\definecolor{codepurple}{rgb}{0.58,0,0.82}
\definecolor{backcolour}{rgb}{0.93,0.93,0.93}
\begin{document}

\def\spacingset#1{\renewcommand{\baselinestretch}%
{#1}\small\normalsize} \spacingset{1}

\if0\blind
{
  \title{\bf Tutorial for Bayesian Factor Models}
  \author[1]{Peter Dunson}
  \author[2]{Ciprian M. Crainiceanu}
  \affil[1]{Department of Mathematics \& Statistics, Kenyon College}
  \affil[2]{Department of Biostatistics, Johns Hopkins University}
  \maketitle
} \fi

\if1\blind
{
  \bigskip
  \bigskip
  \bigskip
  \begin{center}
    {\LARGE\bf Title}
\end{center}
  \medskip
} \fi

\bigskip

\begin{abstract}
Bayesian Factor Models (BFM) are  well-established  models that decompose the observed variability in a set of mean-zero, independent, and uncorrelated factors (random effects). While Factor Analysis (FA) was introduced in 1904 by  \cite{spearman1904}, there has been renewed interest in inferential and computational methods that can adapt to large and complex modern data sets that are now routinely collected in a variety of applications. We provide reproducible, harmonized, and fast software for a variety of recent BFMs that allows the direct comparison of methods and provides a one-stop tutorial for the BFMs and their implementation. We neither endorse nor recommend any of the methods for a particular application; we simply provide a previously unavailable harmonized and reproducible common platform for BFMs. The accompanying \texttt{factorverse} R package is available at \url{https://github.com/peterdunson/factorverse}.
\end{abstract}

\noindent%
\vfill

\newpage
\spacingset{1.75}

\section{Introduction}\label{sec:intro}

Bayesian factor models (BFM) encompass a broad class of methods, with extensions for time series  \citep{aguilar2000,lopes2007}, functional  \citep{matuk2025} and spatial \citep{schmidt2003} data and relaxation of assumptions that include estimation of the latent factor distribution \citep{chandra2023}, mixtures of factor analyzers \citep{ghahramani1996factor,mclachlan2003}, non-Gaussian \citep{murray2013} and/or non-linear \citep{yalcin2001} structure. Despite this generality, the baseline Gaussian linear factor model remains widely used across a range of applications, from genomics to psychology and finance. This paper focuses exclusively on that model, providing the first unified, harmonized, and computationally efficient software platform implementing six modern Bayesian shrinkage priors behind a single consistent interface in the \texttt{factorverse} package. While general-purpose tools such as \texttt{Stan} can be used to fit these models, here we focus on much more efficient dedicated Gibbs samplers developed for BFMs. The paper is developed as a self-contained tutorial that provides a one-stop pragmatic introduction and to modern BFM modeling.

\subsection{Model}\label{subsec:model}
The data structure consists of $P$ measurements $\mathbf{X}_i=(X_{i1},\ldots,X_{iP})^t$ for each individual $i=1,\ldots,N$. A $K$-factor model assumes that
\begin{equation}
X_{ip} = \sum_{k=1}^K \lambda_{pk}b_{ik} + \epsilon_{ip},
\label{eq:basic_model}
\end{equation}
for $i=1,\ldots,N$, and $p=1,\ldots,P$. Here $\lambda_{pk}$ is an unknown ``loading of the factor $k$ on variable $p$'', $b_{ik}$ are unknown ``scores'' for individual $i$ on factor $k$, and $\epsilon_{ip}$ is an error term. The model can be re-written in matrix form as
\begin{equation}
\mathbf{X}_i = \boldsymbol{\Lambda}\,\mathbf{b}_i + \boldsymbol{\epsilon}_i,
\label{eq:basic_model_matrix}
\end{equation}
where $\boldsymbol{\Lambda}$ is the $P\times K$ loading matrix with $\lambda_{pk}$ as the $(p,k)$ entry, $\mathbf{b}_i=(b_{i1},\ldots,b_{iK})^t$ is the $K$-dimensional vector of unknown scores, and $\boldsymbol{\epsilon}_i=(\epsilon_{i1},\ldots,\epsilon_{iP})^t$ is the $P$-dimensional vector of errors. The model assumes that $E(\boldsymbol{b}_i)=\mathbf{0}_K$, the $K\times 1$ dimensional vector of zeros, ${\rm Cov}(\boldsymbol{b}_i)=I_K$, the $K\times K$ dimensional identity matrix, $E(\boldsymbol{\epsilon}_i)=\mathbf{0}_P$, and $\boldsymbol{\Sigma}_\epsilon={\rm Cov}(\boldsymbol{\epsilon}_i)={\rm diag}(\sigma_1^2,\ldots,\sigma^2_P)$, the $P\times P$ diagonal matrix with the vector $(\sigma_1^2,\ldots,\sigma^2_P)$ on the main diagonal and zeros off-diagonal. One typically assumes that $\sigma^2_p>0$ and $K \ll P$, though these are not always explicit assumptions.

These assumptions on the unobserved structure of the data induce assumptions about the observed data. Specifically, $E(\mathbf{X}_i)=\mathbf{0}_P$ and ${\rm Cov}(\mathbf{X}_i)=\boldsymbol{\Omega}=\boldsymbol{\Lambda\Lambda}^t+\boldsymbol{\Sigma}_\epsilon$, which requires the data to be centered (mean zero) and assumes that the covariance matrix can be written as a sum between a low-rank positive semi-definite matrix, $\boldsymbol{\Lambda\Lambda}^t$, and a diagonal matrix, $\boldsymbol{\Sigma}_\epsilon$. In most applications the data are also standardized (variables are centered and divided by their standard deviation) and the assumption is that ${\rm Var}(X_{ip})=1$ for $p=1,\ldots,P$, which translates into ${\rm diag}(\boldsymbol{\Lambda\Lambda}^t)+{\rm diag}(\boldsymbol{\Sigma}_\epsilon)=\mathbf{E}_P$, the $P\times 1$ dimensional vector of ones. Although this is not required, it is a common practice as it simplifies selection of default priors that can be used broadly in Bayesian analyses without incorporating domain knowledge in an application area. The number of observations in the estimated correlation matrix ${\rm Cor}(\mathbf{X}_i)$ is equal to $P\times (P-1)/2$, whereas the number of free parameters in $\boldsymbol{\Lambda\Lambda}^t$ is $PK-\binom{K}{2}$, accounting for the non-identifiability of $\boldsymbol{\Lambda}$ up to a $K\times K$ orthogonal rotation, while the parameters in $\boldsymbol{\Sigma}_\epsilon$ are completely determined by the diagonal terms in $\boldsymbol{\Lambda\Lambda}^t$. Therefore, when $P \gg K$ there are more observations in the sample correlation matrix than parameters in the theoretical correlation matrix. However, the parameters $\boldsymbol{\Lambda}$ and $\mathbf{b}_i$ are not identifiable because for any orthonormal $K\times K$ dimensional  matrix $\mathbf{U}$ the likelihood of model~\eqref{eq:basic_model_matrix} is the same for $(\boldsymbol{\Lambda},\mathbf{b}_i)$ and $(\boldsymbol{\Lambda}\mathbf{U},\mathbf{U}^t\mathbf{b}_i)$. While $\boldsymbol{\Lambda}$ is not identifiable, the product $\boldsymbol{\Lambda}\boldsymbol{\Lambda}^t$ is.

\subsection{Connections with linear regression and PCA}\label{subsec:connections}
Factor analysis models~\eqref{eq:basic_model}--\eqref{eq:basic_model_matrix} differ from linear regression in that the covariates $\mathbf{b}_i$ are unknown and must be estimated from the data, with the loadings $\boldsymbol{\Lambda}$ playing the role of regression coefficients on these latent covariates. Factor models are also different from Principal Component Analysis models because: (1) the matrix $\boldsymbol{\Lambda}$ is not required to be orthonormal and its columns are not required to explain the maximum residual variability; (2) the covariance matrix ${\rm Cov}(\mathbf{b}_i)=I_K$, while the covariance of the PC scores is diagonal with different variances (these could be absorbed in $\boldsymbol{\Lambda}$); and (3) the error term $\boldsymbol{\epsilon}_i$ is often left unspecified in PCA or error variances are assumed equal. Let $\boldsymbol{\Lambda}=\mathbf{U}_K\mathbf{D}_K\mathbf{V}_K^t$ be the rank-K  singular value decomposition (or compact SVD) of $\boldsymbol{\Lambda}$, where $\mathbf{U}_K$ is the $P\times K$ dimensional matrix such that $\mathbf{U}_K^t\mathbf{U}_K=I_K$ (columns of $\mathbf{U}$ are orthonormal), $\mathbf{D}_K={\rm diag}(d_1,\ldots,d_K)$ is the $K\times K$ diagonal matrix of singular values, and $\mathbf{V}_K^t$ is the $K\times K$ dimensional matrix such that $\mathbf{V}_K^t\mathbf{V}_K=I_K$. If we define $\widetilde{\boldsymbol{\Lambda}}=\mathbf{U}_K$ and $\widetilde{\mathbf{b}}_i=\mathbf{D}_K\mathbf{V}_K^t\mathbf{b}_i$, we have $\widetilde{\boldsymbol{\Lambda}}\widetilde{\mathbf{b}}_i=\boldsymbol{\Lambda}\mathbf{b}_i$, $\widetilde{\boldsymbol{\Lambda}}^t\widetilde{\boldsymbol{\Lambda}}=I_K$, and ${\rm Cov}(\widetilde{\mathbf{b}}_i)=\mathbf{D}_K\mathbf{V}_K^t\mathbf{V}_K\mathbf{D}_K={\rm diag}(d_1^2,\ldots,d_K^2)$, which is the structure of the PCA. A crucial difference is that PCA/SVD makes no assumption about the correlation of the residuals $\mathbf{X}_i-\mathbf{\Lambda}\mathbf{b}_i$, whereas the Gaussian linear factor model considered here assumes that this correlation is diagonal and the variance of the errors ($\sigma^2_p$, $p=1,\ldots,P$) are strictly positive. 

\subsection{Building up the intuition using the \texorpdfstring{$1$}{1}-factor model}\label{subsec:intuition}  

Consider the case when there are $P=3$ variables and assume that they are centered and standardized. A one-factor (K=1) model has the following form
$$X_{ip}=\lambda_pb_i+\epsilon_{ip},$$
where $E(b_i)=0$, ${\rm Var}(b_i)=1$, $E(\epsilon_{ip})=0$, ${\rm Var}(\epsilon_{ip})=\sigma^2_p$, and $b_i,\epsilon_{i1},\epsilon_{i2}, \epsilon_{i3}$ are mutually uncorrelated among themselves and across study participants, $i$. The idea is that there is some underlying personal characteristic (random effect, latent factor), $b_i$, that influences the measurement $X_{ip}$ through the population-level loading $\lambda_p$, which characterizes the impact of the factor on the $p$th observed variable. The assumption is that the unexplained subject-specific residuals $\epsilon_{ip}=X_{ip}-\lambda_pb_i$ are not correlated with the personal characteristics, $b_i$, or among themselves (across measurements $p=1,2,3$). This implies that $1={\rm Var}(X_{ip})=\lambda_p^2+\sigma^2_p$ and $\rho_{pq}={\rm Cor}(X_{ip},X_{iq})=\lambda_p\lambda_q$ for any $p\neq q$. Therefore, $\sigma^2_p=1-\lambda_p^2$ and $\lambda_p^2=\frac{\rho_{pq}\rho_{pr}}{\rho_{qr}}$ for $p\neq q \neq r\neq p$. This indicates that $\lambda_p=\sqrt{\frac{\rho_{pq}\rho_{pr}}{\rho_{qr}}}$ or $\lambda_p=-\sqrt{\frac{\rho_{pq}\rho_{pr}}{\rho_{qr}}}$ and requires that $\rho_{pq}\neq 0$ and $\frac{\rho_{pq}\rho_{pr}}{\rho_{qr}}>0$ for all $p\neq q \neq r\neq p$; we call these compatibility conditions for the $K=1$ factor model with the data $(X_{i1},X_{i2},X_{i3})^t$ for $i=1,\ldots,N$ to differentiate them from the identifiability conditions. This provides  simple methods of moments (MoM) estimators for $\lambda_p^2$ and $\sigma^2_p=1-\lambda^2_p$ and shows under which assumptions the model is identifiable. In this example, identifiability is achieved by choosing the sign of $\lambda_1$ (recall the identifiability up to the orthonormal matrix $\mathbf{U}$ discussed in section~\ref{subsec:connections}). Indeed, if we choose $\lambda_1>0$  then $\lambda_2$ has the sign of $\rho_{12}$ and $\lambda_3$ has the sign of $\rho_{13}$.

\section{Bayesian Factor Analysis Review}\label{subsec:litreview}

Factor analysis originated in psychometrics with \citet{spearman1904}. Numerous estimation strategies have since been developed. Classical approaches include maximum likelihood estimation \citep{lawley1962}, often carried out via the EM algorithm \citep{rubin1982}, and generalized method of moments estimation \citep{bollen2014}. Other frequentist procedures include principal-factor and least-squares methods \citep{joreskog1967} and optimization techniques from structural equation modeling \citep{bollen1989}. Bayesian approaches range from early formulations \citep{martin1975,press1989} to modern MCMC methods \citep{lopes2004}, dynamic and nonparametric extensions \citep{aguilar2000,lopes2007, chandra2023}, and structured shrinkage priors \citep{carvalho2010,bhattacharya2011,george1993,rockova2018,bhattacharya2015}. Scalable alternatives include variational inference for factor analyzers \citep{ghahramani1996factor} and related VI methods \citep{blei2017}.

A close inspection of model~\eqref{eq:basic_model}, reveals that it contains many random effects and that the full conditionals of $\lambda_{pk}$ and $b_{ik}$ are easy to obtain because the model is linear when one conditions on the other parameters. This provides an ideal situation for Bayesian models that can efficiently explore the posterior distribution using Gibbs sampling. This observation combined with the explosion in the number and complexity of new data, led to a renewed interest in Bayesian models for factor analysis. The work of \citet{geweke1996} provided a landmark contribution, introducing an efficient Gibbs sampler that implemented sign and ordering constraints on the loadings to obtain a coherent and computationally stable parametrization for a fixed number of factors $K$. This computational breakthrough combined with subsequent work on inference on the number of factors $K$ \citep{lopes2004} and the ever-improving Bayesian computation techniques, set the stage for important developments in Bayesian factor models. Earlier approaches enforced identifiability by constraining $\boldsymbol{\Lambda}$ to be lower triangular with positive diagonal entries, though this introduces an undesirable dependence on variable ordering. The most common modern strategy is to run posterior computation in the over-parameterized space and restrict inference to rotation-invariant functionals such as $\boldsymbol{\Lambda\Lambda}^t$ and $\boldsymbol{\Omega}$, which mix well without constraints; if inference on the loadings or factors themselves is required, post-processing to rotationally align MCMC samples is needed \citep{poworoznek2025}. Selecting the number of factors $K$ has also evolved considerably; the reversible jump MCMC approach of \citet{lopes2004} required repeated refitting and is rarely used today, with modern practice instead relying on over-fitted factor models that shrink unnecessary columns of $\boldsymbol{\Lambda}$ toward zero, or on refitting for a range of $K$ values and selecting via WAIC, which performs well in latent factor settings.

A central theme in the subsequent two decades has been to induce \textbf{shrinkage} in the factor loading matrix, $\boldsymbol{\Lambda}$, using priors on the entries of this matrix. In principle, this serves two purposes: enhancing interpretability and estimating the number of factors $K$ directly from the data. Several approaches have been proposed including the multiplicative gamma process shrinkage (MGPS) prior \citep{bhattacharya2011}, the Dirichlet-Laplace (DL) prior \citep{bhattacharya2015}, the spike-and-slab priors \citep{rockova2018}, the horseshoe prior \citep{carvalho2010}, Bayesian Analysis with Structured Sparsity \citep{zhao2016}, and Mass-Nonlocal score \citep{huang2025}. We provide the main features of each approach and then provide harmonized software for their implementation.

\section{Implementation of Methods}\label{sec:implement}

\subsection{Data structure and priors}\label{subsec:data_structure} The data consists of an $N\times P$ dimensional matrix, where each row corresponds to a subject (experimental unit) and each column corresponds to a specific measurement. Missing observations under missing at random (MAR) assumptions are straightforward to handle by allowing only the observed variables for each subject to contribute to the relevant full conditional distributions; here we assume no missing data for simplicity. Data can also be stored in long format, where the first column contains the observations $X_{ip}$ and the second column contains the subject identifier variable. The additional assumptions are that data are properly normalized (every variable has mean $0$ and variance $1$) and every variable has a marginal Normal distribution. 

Bayesian inference under model~\eqref{eq:basic_model} shares the same likelihood across all methods considered here, with differences arising entirely from the choice of prior on $\boldsymbol{\Lambda}$. The vast majority of models assume that $\mathbf{b}_i\sim N(0,I_K)$, a standard $K$-variate normal and that $\boldsymbol{\epsilon}_i\sim N(0,\boldsymbol{\Sigma}_\epsilon)$, where $\boldsymbol{\Sigma}_\epsilon$ is a $P\times P$ dimensional diagonal matrix with $\sigma^2_p$ as the $p$ element on the diagonal. Residual precision parameters follow independent Gamma priors $\sigma^{-2}_p\sim {\rm Gamma}(a_s,b_s)$, where $a_s=1$ and $b_s=0.3$. In this paper we use the ${\rm Gamma}(a,b)$ parameterization, where $a$ is the shape parameter and $b$ is the rate parameter with the mean of the distribution equal to $a/b$ and variance $a/b^2$. We now review some of the major priors proposed in the literature for $\boldsymbol{\Lambda}$.

\subsection{Multiplicative Gamma Process Shrinkage Prior (MGPS)}\label{subsec:MGPS}

The MGPS prior was designed for over-fitted factor models, where the number of factors is initialized at a generous upper bound and unnecessary factors are removed by shrinkage rather than by model selection. It achieves this through a multiplicative structure on the column-wise shrinkage that increases with the factor index, so that later columns of $\boldsymbol{\Lambda}$ are shrunk more aggressively toward zero, effectively deleting superfluous factors while retaining conjugate Gibbs updates throughout.

The multiplicative gamma process shrinkage prior \citep{bhattacharya2011} induces adaptive, column-wise shrinkage on the loading matrix $\boldsymbol{\Lambda}$. Each loading entry $\lambda_{pk}\sim N(0, \sigma^2_{\lambda,pk})$ where $\sigma^2_{\lambda,pk} = (\phi_{pk}\tau_{k})^{-1}$, with $\phi_{pk}\sim\mathrm{Gamma}(\tfrac{\nu}{2},\tfrac{\nu}{2})$ and $\tau_{k}=\prod_{l=1}^{k}\delta_l$ with $\delta_{1}\sim\mathrm{Gamma}(a_{1},b_{1})$ and $\delta_{l}\sim\mathrm{Gamma}(a_{2},b_{2})$ for $l>1$.  In our implementation we set $\nu=3$, $a_{1}=2.1$, $b_{1}=1.0$, $a_{2}=3.1$, $b_{2}=1.0$, as suggested by \citet{durante2017}.

Posterior inference is implemented using the Gibbs sampler in \texttt{C++} using the \texttt{RcppArmadillo} \citep{armmanual,armpackage} package in \texttt{R} \citep{Rmanual}. We use the full conditionals provided in \citep{bhattacharya2011} (page 296) and include them here for presentation completeness. Notation is slightly changed and streamlined to ensure harmonization across papers and methods. The full conditionals are

\begin{itemize}
\item[1.] $[\mathbf{b}_{i}|{\rm others}]\sim N\{(I_{K}+\boldsymbol{\Lambda}^{\!\top}\boldsymbol{\Sigma}_\epsilon^{-1}\boldsymbol{\Lambda})^{-1}\boldsymbol{\Lambda}^{\!\top}\boldsymbol{\Sigma}_\epsilon^{-1}\mathbf{X}_{i},(I_{K}+\boldsymbol{\Lambda}^{\!\top}\boldsymbol{\Sigma}_\epsilon^{-1}\boldsymbol{\Lambda})^{-1}\}$
\item[2.] $[\boldsymbol{\lambda}_{p\cdot}|{\rm others}]\sim N[\boldsymbol{\Sigma}_{\lambda,p}^{-1}\{\sigma^{-2}_p\sum_{i}\mathbf{b}_{i}X_{ip}\},\boldsymbol{\Sigma}_{\lambda,p}^{-1}],$\\ where $\boldsymbol{\Sigma}_{\lambda,p}=\sigma^{-2}_{p}\sum_{i}\mathbf{b}_{i}\mathbf{b}_{i}^t+\mathrm{diag}(\boldsymbol{\phi}_{p\cdot}\boldsymbol{\tau})$, $\boldsymbol{\lambda}_{p\cdot}$ is the $1\times K$ dimensional row $p$ vector of  $\boldsymbol{\Lambda}$, and $\mathrm{diag}(\boldsymbol{\phi}_{p\cdot}\boldsymbol{\tau})=\mathrm{diag}(\phi_{p1}\tau_1,\ldots,\phi_{pK}\tau_K)$ is a $K\times K$ dimensional diagonal matrix.
\item[3.] $[\sigma_{p}^{-2}|{\rm others}]\sim {\rm Gamma}(a_{s}+n/2,b_{s}+\tfrac12\sum_{i}(X_{ip}-\mathbf{b}_{i}^t\boldsymbol{\lambda}_{p\cdot})^{2})$
\item[4.] $[\phi_{pk}|{\rm others}] \sim \mathrm{Gamma}(\tfrac{\nu+1}{2},\tfrac{\nu+\tau_{k}\lambda_{pk}^{2}}{2})$
\item[5.] $[\delta_{k}|{\rm others}]\sim \mathrm{Gamma}(a_{k}+\tfrac{P}{2}(K-k+1),b_{k}+\tfrac12\sum_{l\geq k}\tau_l^{(k)}\sum_{p}\phi_{pl}\lambda_{pl}^{2})$, where $(a_{k},b_{k})=(a_{1},b_{1})$ if $k=1$ and $(a_{2},b_{2})$ otherwise, and $\tau_l^{(k)}=\prod_{t\neq k}\delta_t$.
\end{itemize}

All the full conditionals are easy to simulate from. In Section~\ref{subsubsec:mgps_post} we describe how to implement this model in \texttt{C++}. The advantage of the \texttt{C++} implementation is that the entire Gibbs sampler runs in compiled code, avoiding the per-iteration overhead of the \texttt{R} interpreter. Because factor-model sampling requires many thousands of iterations, each involving dense matrix operations (the $K\times K$ solve in the loading and score updates, repeated across $P$ variables and $n$ subjects), this overhead dominates runtime in a pure \texttt{R} implementation. Routing the linear algebra through \texttt{RcppArmadillo} keeps these operations in optimized \texttt{C++} and yields substantial speedups, which matters in the simulation studies of Section~\ref{sec:simulation}, where every method is fit across many replicates and parameter settings.

\subsection{Spike-and-Slab LASSO Prior (SSL)}\label{subsec:SSL}

The SSL prior was designed to improve on continuous shrinkage priors by introducing a discrete mixture structure that more cleanly separates zero and nonzero loadings. Rather than shrinking all loadings toward zero by a continuous penalty, it assigns each loading to either a tight spike or a loose slab Laplace distribution, allowing the data to determine which loadings are effectively zero and which are active, while retaining conjugate Gibbs updates through a scale mixture of normals representation.

The main difference between the Spike-and-Slab LASSO prior \citep{rockova2018} and the MGPS prior is the prior distribution on $\boldsymbol{\Lambda}$. More precisely, the Spike-and-Slab LASSO prior is a mixture of two Laplace distributions, one tight (spike) and one loose (slab) around zero \citep{ishwaran2005}. Specifically the distribution of $\lambda_{pk}$ is induced by the conditional distributions
$$[\lambda_{pk}\mid z_{pk}]=(a_{z_{pk}}/2)\exp(-a_{z_{pk}}|\lambda_{pk}|),$$
where $a_{0}\gg a_{1}$ and $z_{pk}\sim {\rm Bernoulli}(\theta)$ is a latent Bernoulli variable. Thus, the marginal distribution is $\lambda_{pk}\sim (1-\theta)(a_{0}/2)\exp(-a_{0}|\lambda_{pk}|)+\theta(a_{1}/2)\exp(-a_{1}|\lambda_{pk}|)$. We set $a_{0}=20$ and $a_{1}=0.2$, and place a $\mathrm{Beta}(1,P)$ prior on $\theta$ so that the expected fraction of nonzero loadings adapts to the data rather than being fixed in advance.

The main change from the MGPS full conditionals is sampling the conditionally independent indicators $z_{pk}$ from the normalized probability \newline $\Pr(z_{pk}=1\mid\boldsymbol{\Lambda}) = \frac{\theta a_{1} \exp(-a_{1}|\lambda_{pk}|)}{\theta a_{1} \exp(-a_{1}|\lambda_{pk}|) + (1-\theta) a_{0} \exp(-a_{0}|\lambda_{pk}|)}$, then sampling an auxiliary scale $w_{pk}$ that represents the Laplace prior as a scale mixture of normals. The data augmentation with the unobserved variables $w_{pk}$ replaces sampling from a complex full conditional with two simple ones: $\lambda_{pk}\mid w_{pk}\sim N(0,w_{pk})$, and $w_{pk}$ drawn from its generalized inverse Gaussian full conditional. The reciprocal $1/w_{pk}$ serves as the prior precision for the loading, replacing the MGPS precision $\mathrm{diag}(\boldsymbol{\phi}_{p\cdot}\boldsymbol{\tau})$ in the loadings update, while the factor score and residual precision draws proceed identically to the MGPS method.

\subsection{Dirichlet--Laplace Prior (DL)}\label{subsec:DL}

The DL prior was designed to achieve more aggressive and theoretically grounded shrinkage than the MGPS prior by combining a global-local scale structure with a Dirichlet allocation of shrinkage across factors. The Dirichlet component ensures that each row of $\boldsymbol{\Lambda}$ receives the same total amount of shrinkage, partitioned adaptively across factors, while the Laplace tails shrink small loadings more heavily than a Gaussian prior while protecting large ones from overshrinkage.

The Dirichlet--Laplace prior of \cite{bhattacharya2015} replaces the independent normal priors $\lambda_{pk}\sim N(0,(\phi_{pk}\tau_k)^{-1})$ of \cite{bhattacharya2011} with independent Laplace priors $\lambda_{pk}\sim\mathrm{Laplace}(\phi_{pk}\tau_p)$, where the prior is applied independently to each row $\boldsymbol{\lambda}_{p\cdot}$ with $\boldsymbol{\phi}_{p\cdot}\sim\mathrm{Dirichlet}(a,\dots,a)$ and \newline $\tau_p\sim\mathrm{Gamma}(Ka,1/2)$. The Dirichlet--Laplace prior is used for the same reasons as the Normal prior, though there are also important philosophical and technical differences. First, the Laplace prior has thicker tails than the Normal, which shrinks less large effects and more small effects. Second, the shrinkage parameters are $\phi_{pk}\tau_p$, where $\sum_{k=1}^K \phi_{pk}=1$ because $\boldsymbol{\phi}_{p\cdot}=(\phi_{p1},\ldots,\phi_{pK})^t$ has a Dirichlet distribution and $\tau_p$ represents the row-specific scale. Therefore, each row of $\boldsymbol{\Lambda}$ is assigned its own total amount of smoothing, $\tau_p$. The Dirichlet prior on the $K\times 1$ dimensional vector $\boldsymbol{\phi}_{p\cdot}$ thus acts as a partition of the row-total shrinkage $\tau_p$ into its components. We initialize $a=1/P$. This row-specific construction follows the original Dirichlet--Laplace formulation of \cite{bhattacharya2015} and the column-wise Dirichlet--Laplace used in \cite{ferrari2021}; each $\tau_p\sim\mathrm{Gamma}(Ka,1/2)$.

Posterior inference uses the  Gibbs sampler for MGPS prior introduced in Section~\ref{subsec:MGPS}, with some changes. First, the full conditional for the vector $[\boldsymbol{\phi}_{p\cdot}|{\rm others}]$ is obtained by sampling independently $T_{pk}\sim{\rm GIG}(a-1,2|\lambda_{pk}|, 1)$ and setting $\phi_{pk}=T_{pk}/\sum_{l=1}^K T_{pl}$. Second, the full conditionals for $[\lambda_{pk}|{\rm others}]$ and $[\tau_p|{\rm others}]$ depend on the auxiliary variables $\psi_{pk}$. To ensure conjugacy, the following scale mixture of Normals representation is used: $\lambda_{pk}\mid \psi_{pk} \sim N(0, \sigma^2_{\lambda,pk})$ where $\sigma^2_{\lambda,pk} = \psi_{pk}\phi_{pk}^2\tau_p^2$ with $\psi_{pk} \sim \mathrm{Exp}(1/2)$; see Section 2.4 in \cite{bhattacharya2015} for this computational insight.

Based on this augmentation, the full conditionals are: $[\tau_p|{\rm others}]\sim {\rm GIG}(K(a-1),2\sum_{k}|\lambda_{pk}|/\phi_{pk}, 1)$ for each row $p$, and the auxiliary variables $[\psi_{pk}^{-1}|{\rm others}]\sim {\rm IGauss}(\phi_{pk}\tau_p/|\lambda_{pk}|,1)$. Here the distribution ${\rm GIG}(a,b,c)$ is the generalized inverse Gaussian distribution with probability density function $f(x|a,b,c)\propto x^{a-1}\exp\{-\frac{1}{2}(b/x+cx)\}$. Note that the inverse-Gaussian distribution corresponds to ${\rm IGauss}(\mu,\lambda)={\rm GIG}(-1/2,\lambda, \lambda/\mu^2)$ in this notation.

\subsection{Horseshoe Prior (HS)}\label{subsec:HS}

The horseshoe prior was designed to achieve aggressive shrinkage of near-zero loadings while placing essentially no shrinkage on large ones, a property that distinguishes it from both the Gaussian priors of MGPS and the Laplace priors of DL and SSL. This behavior arises from the heavy-tailed half-Cauchy distributions on the local and global scales, which allow the posterior to concentrate tightly around zero for irrelevant loadings while leaving large signals nearly unshrunk.

The horseshoe prior of \cite{carvalho2010} replaces the MGPS/SSL prior on the loading matrix $\boldsymbol{\Lambda}$ by imposing global--local shrinkage: each loading entry $\lambda_{pk}\mid\tau^{2},\phi_{pk}^{2}\sim N(0,\sigma^2_{\lambda,pk})$ where $\sigma^2_{\lambda,pk} = \tau^{2}\phi_{pk}^{2}$ with local scales $\phi_{pk}\sim C^{+}(0,1)$ and a global scale $\tau\sim C^{+}(0,1)$, where $C^{+}(0,1)$ denotes the positive half-Cauchy distribution with parameters $(0,1)$. Just as in the case of the DL prior, a data augmentation approach can be used to obtain conjugate Gibbs sampling for the HS prior. Here we will use the data augmentation approach of \cite{makalic2016}, who  introduced the auxiliary variables $\nu_{pk}$ and $\xi$ so that the full conditional updates are $\phi_{pk}^{2}\sim \mathrm{IG}\bigl(1,\,\lambda_{pk}^{2}/(2\tau^{2})+1/\nu_{pk}\bigr)$, $\nu_{pk}\sim \mathrm{IG}(1,\,1+1/\phi_{pk}^{2})$, $\tau^{2}\sim \mathrm{IG}\bigl((PK+1)/2,\,\sum_{p,k}\lambda_{pk}^{2}/(2\phi_{pk}^{2})+1/\xi\bigr)$ and $\xi\sim \mathrm{IG}(1,\,1+1/\tau^{2})$. Here $\mathrm{IG}(a,b)$ denotes the inverse-Gamma distribution with shape $a$ and rate $b$, following \citet{makalic2016}. All parameters are sampled exactly as in \S\ref{subsec:MGPS} with the exception of the full conditionals  corresponding to $\lambda_{pk}$, which are replaced by these four full conditionals.

\subsection{Bayesian Group Factor Analysis with Structured Sparsity (BASS)}\label{subsec:BASS}

BASS extends the global-local shrinkage framework by introducing a deeper hierarchical structure that operates at both the variable and factor level, and additionally incorporates a binary indicator for each factor that switches between sparse and dense shrinkage regimes. This design allows the model to adapt not just to the sparsity of individual loadings but to the overall structure of each factor, making it more flexible than priors that apply a single shrinkage mechanism uniformly across all entries of $\boldsymbol{\Lambda}$.

The BASS prior of \citet{zhao2016} starts with independent normal priors on the factor loadings $\lambda_{pk}\sim N(0,\sigma^2_{\lambda,pk})$ where $\sigma^2_{\lambda,pk} = \theta_{pk}^{-1}$, with $\theta_{pk}$ denoting the prior precision. The method imposes a three-level hierarchical Gamma prior as follows $\theta_{pk}\sim\mathrm{Gamma}(a,\delta_{pk})$,  $\delta_{pk}\sim\mathrm{Gamma}(b,\omega_k)$, $\omega_k\sim\mathrm{Gamma}(c,\rho_k)$,  $\rho_k\sim\mathrm{Gamma}(d,\eta)$, $\eta\sim\mathrm{Gamma}(e,\gamma)$, and $\gamma\sim\mathrm{Gamma}(f,\nu)$. We set $a=b=c=d=e=f=0.5$ and $\nu=1$, which corresponds to the horseshoe prior at each level of the hierarchy.

Each factor $k$ is additionally assigned a binary indicator $z_k\sim\mathrm{Bernoulli}(\pi)$ that selects between a \emph{sparse} regime ($z_k=1$), in which the local precisions $\theta_{pk}$ are updated element-wise, and a \emph{dense} regime ($z_k=0$), in which all loadings in factor $k$ share the common scale $\omega_k$. The mixture proportion $\pi$ is assigned a $\mathrm{Beta}(1,1)$ prior and updated from its conjugate full conditional $\pi\mid\mathbf{z}\sim\mathrm{Beta}(1+\sum_k z_k,\,1+K-\sum_k z_k)$.

This prior set up leads to simple Gamma full conditional updates for the precision parameters: $\theta_{pk}\sim\mathrm{Gamma}(a+\tfrac12,\,\delta_{pk}+\tfrac12\lambda_{pk}^2)$, $\delta_{pk}\sim\mathrm{Gamma}(b+1,\,\omega_k+\theta_{pk})$, $\omega_k\sim\mathrm{Gamma}(c+P,\,\rho_k+\sum_p\delta_{pk})$, $\rho_k\sim\mathrm{Gamma}(d+1,\,\eta+\omega_k)$, $\eta\sim\mathrm{Gamma}(e+K,\,\gamma+\sum_k\rho_k)$, and $\gamma\sim\mathrm{Gamma}(f+1,\,\nu+\eta)$. The factor-score and residual-precision updates are identical to those in Section~\ref{subsec:MGPS}, while the loadings update takes the same Gaussian form with $\mathrm{diag}(\boldsymbol{\theta}_{p\cdot})$ as the prior precision.

\subsection{Mass-Nonlocal Score Prior (MNL)}\label{subsec:MNL}

The MNL prior takes a fundamentally different approach from the other methods considered here: rather than placing a shrinkage prior on the loadings $\boldsymbol{\Lambda}$, it modifies the distribution of the factor scores $\mathbf{b}_i$ directly. It was designed for settings where many subjects are expected to have a true score of exactly zero on a given factor, using a spike-and-slab mixture on the scores with a product-moment nonlocal slab that places zero density at the origin, actively separating nonzero scores from zero rather than merely shrinking them.

The mass-nonlocal score prior of \citet{huang2025} modifies the factor score distribution by introducing a spike-and-product-moment (pMOM) mixture in place of the standard normal assumption, while retaining the MGPS prior on the loadings. Each score entry $b_{ik}$ is endowed with a Bernoulli latent indicator $Z_{ik}\sim\mathrm{Bernoulli}(\theta_{k})$ so that $b_{ik}=0$ when $Z_{ik}=0$, and when $Z_{ik}=1$ its slab follows a pMOM density, a product-moment prior whose density is proportional to $b_{ik}^{2}$ times a Gaussian and which therefore vanishes at the origin, separating nonzero scores away from zero. The proportion of nonzero scores $\theta_{k}$ is assigned a $\mathrm{Beta}(1,5)$ prior and updated from its conjugate full conditional; the pMOM dispersion $\psi_{k}$ is not sampled but set deterministically as $\psi_{k}=1/(3\theta_{k})$, following the reference implementation.

Posterior inference reuses the Gibbs sampler of Section~\ref{subsec:MGPS}, with the following score-specific steps inserted in place of the standard Gaussian draws. First, each indicator $Z_{ik}$ is sampled from its full conditional, which marginalizes the score $b_{ik}$ in closed form (Section~\ref{subsubsec:MNL_post}). Next, each $\theta_{k}$ is drawn from its Beta full conditional from the slab counts in column $k$ of $\mathbf{Z}$, and $\psi_{k}$ is recomputed. Nonzero scores are then drawn by a short random-walk Metropolis--Hastings chain targeting the score full conditional, since the pMOM slab combined with the Gaussian likelihood does not yield a standard form; scores with $Z_{ik}=0$ are set exactly to zero. Finally, after updating $\mathbf{b}$, the usual MGPS updates for $\boldsymbol{\Lambda}$, the local shrinkage and global multipliers, and the residual precisions proceed exactly as in Section~\ref{subsec:MGPS}. Following the reference, posterior summaries for $\boldsymbol{\Lambda}$ and $\boldsymbol{b}$ use the median over draws.

\section{Implementation in the \texttt{factorverse} Package}\label{sec:code}
As shown in Section~\ref{sec:implement}, the various approaches share most of their full conditionals, differing chiefly in the prior placed on $\boldsymbol{\Lambda}$ (or, for MNL, on the factor scores $\mathbf{b}$). The methods are platform agnostic and could be implemented in any software. Here we use our own \texttt{C++} implementation via \texttt{RcppArmadillo}. The wrapper function exposes a single consistent interface across all methods. Below we provide the computational infrastructure for the function, but here we show the user interface:

\begin{minted}[mathescape, baselinestretch=1.2, bgcolor=LightGray, fontsize=\footnotesize]{R}
fit <- fit_bfm(method = "mnl", dat = data, k0 = 5,
                      nrun = 16000, burn = 8000, chains = 4)
\end{minted}

The \texttt{fit} object is a list containing the posterior simulations for all model parameters, returned in the \texttt{posterior} element (the loadings $\boldsymbol{\Lambda}$, scores $\mathbf{b}$, residual variances $\boldsymbol{\Sigma}$, and per-iteration factor count), along with the posterior mean loadings in \texttt{Lambda\_hat}. A core design principle of \texttt{factorverse} is its consistent user interface with a standardized syntax. This allows the user to fit different models, extract posterior summaries, and compare results across models, both in simulations and on real datasets.

\subsection{C++ Implementation}\label{subsec:c++}
We show first the implementation of the Bayesian factor model with the MGPS prior and start with the common conditional distributions followed by the MGPS-specific updates (for $\lambda_{pk}$). In subsequent sections we provide the details required for the updates of the $\lambda_{pk}$ parameters for the other methods, as described in Section~\ref{sec:implement}. 

\begin{table}[ht]
\centering
\caption{Notation mapping between statistical formulation and C++ implementation.}
\label{tab:notation}
\footnotesize
\begin{tabular}{lll}
\toprule
Parameter & Statistical & C++ \\
\midrule
Factor scores & $\boldsymbol{b}_i$ & \texttt{b} \\
Loadings & $\boldsymbol{\Lambda}$ & \texttt{lambda} \\
Residual precisions & $\sigma_p^{-2}$ & \texttt{ps} \\
Posterior covariance & $\mathbf{V}_b$ & \texttt{Vb} \\
Posterior precision & $\mathbf{V}_b^{-1}$ & \texttt{V} \\
Prior precision (loadings) & $\phi_{pk}\tau_k$ & \texttt{Plam} \\
Local shrinkage & $\phi_{pk}$ & \texttt{psipk} \\
Global multipliers & $\delta_k$ & \texttt{delta} \\
Cumulative shrinkage & $\tau_k$ & \texttt{tauk} \\
Degrees of freedom & $\nu$ & \texttt{df} \\
Data matrix & $\mathbf{X}$ & \texttt{X} \\
Number of factors & $K$ & \texttt{K} \\
Sample size & $n$ & \texttt{n} \\
Number of variables & $P$ & \texttt{P} \\
\bottomrule
\end{tabular}
\end{table}

\FloatBarrier

\subsubsection{Common full conditionals}\label{subsubsec:likelihood}

Up to the method-specific augmentation and shrinkage parameters, the full conditionals for the factor scores ($\boldsymbol{b}$), loadings ($\boldsymbol{\Lambda}$), and residual precisions ($\sigma_p^{-2}$) share a common form; we refer to these as the common full conditionals. The residual precision update is identical across all methods considered here, since they share the same Gamma prior on $\sigma_p^{-2}$. The loadings update has the same Gaussian form throughout, with only the prior precision matrix swapped out for the method-specific choice. The factor-score update is shared by every method except MNL, which models the scores with a non-Gaussian prior and therefore has a different score full conditional.

{\it Sampling Factor Scores.}
The conditional distribution of the factor scores $\boldsymbol{b}_i$ is the $K$-variate Gaussian  $N_K(\mathbf{V}_b\boldsymbol{\Lambda}^t\boldsymbol{\Sigma}_\epsilon^{-1}\mathbf{x}_i, \mathbf{V}_b)$ where the posterior precision is $\mathbf{V}_b^{-1} = \mathbf{I}_K + \boldsymbol{\Lambda}^t\boldsymbol{\Sigma}_\epsilon^{-1}\boldsymbol{\Lambda}$. This is implemented below as follows

\begin{minted}[mathescape, baselinestretch=1.2, bgcolor=LightGray, fontsize=\footnotesize]{cpp}
// Compute Lambda * diag(psi) 
arma::mat Lps = lambda.each_col() % ps;
// Compute the posterior precision matrix V
arma::mat V = arma::eye<arma::mat>(K,K) + Lps.t() * lambda;

// Force symmetry and add small diagonal term for stability
V = 0.5*(V + V.t());
V.diag() += 1e-8;

// Use Cholesky decomposition to compute the covariance (inverse of V):
// Cholesky: V = LL', so V_b = (L')^{-1}L^{-1}
arma::mat S = arma::inv(arma::trimatu(arma::chol(V)));
arma::mat Vb = S * S.t();  // posterior covariance

// Sample scores for all subjects and factors simultaneously:
return X * Lps * Vb + arma::randn<arma::mat>(n,K) * S.t();
\end{minted}

{\it Sampling Factor Loadings.}
Recall that the distribution of the $K$-dimensional vector of loadings $\boldsymbol{\lambda}_{p\cdot}=(\lambda_{p1},\ldots,\lambda_{pK})$ (row $p$ in the $\boldsymbol{\Lambda}$ matrix) has the conditional distribution $[\boldsymbol{\lambda}_{p\cdot}|{\rm others}]\sim N[\boldsymbol{\Sigma}_{\lambda,p}^{-1}\{\sigma^{-2}_p\sum_{i}\mathbf{b}_{i}X_{ip}\},\boldsymbol{\Sigma}_{\lambda,p}^{-1}]$, where $\boldsymbol{\Sigma}_{\lambda,p}=\sigma^{-2}_{p}\sum_{i}\mathbf{b}_{i}\mathbf{b}_{i}^t+\mathrm{diag}(\boldsymbol{\phi}_{p\cdot}\boldsymbol{\tau})$ and $\mathrm{diag}(\boldsymbol{\phi}_{p\cdot}\boldsymbol{\tau})=\mathrm{diag}(\phi_{p1}\tau_1,\ldots,\phi_{pK}\tau_K)$ is a $K\times K$ dimensional diagonal matrix. Denote by $\boldsymbol{M}_p=\boldsymbol{\Sigma}_{\lambda,p}$ and by $\boldsymbol{m}_p=\sigma^{-2}_p\sum_{i}\mathbf{b}_{i}X_{ip}$ the full conditional distribution can be written as $[\boldsymbol{\lambda}_{p\cdot}|{\rm others}]\sim N(\boldsymbol{M}_p^{-1}\boldsymbol{m}_p,\boldsymbol{M}_p^{-1})$. Note that $\boldsymbol{M}_p=\boldsymbol{\Sigma}_{\lambda,p}$ contains $\sum_{i}\mathbf{b}_{i}\mathbf{b}_{i}^t$, which needs to be computed only once. The \texttt{C++} code is then 

\begin{minted}[mathescape, baselinestretch=1.2, bgcolor=LightGray, fontsize=\footnotesize]{cpp}
arma::mat b2 = b.t() * b;  // K x K, shared across all p

// Loop through each variable and sample its loading vector:
for(int p=0; p<P; ++p) {
    // M_p = diag(prior precision) + psi_p * b'b
    arma::mat Mp = arma::diagmat(Plam.row(p).t()) + ps(p) * b2;

    // Symmetrize and stabilize matrix before inverting
    Mp = 0.5*(Mp + Mp.t());
    Mp.diag() += 1e-8;

    // Calculate the inverse of Mp = LL' using the Cholesky decomposition: 
    arma::mat Linv = arma::inv(arma::trimatu(arma::chol(Mp)));
    
    // Mean component: psi_p * b' * x_p
    arma::vec mp = ps(p) * b.t() * X.col(p);

    // Sample lambda_p from N(M_p^{-1}m_p, M_p^{-1}) 
    out.row(p) = (Linv*arma::randn<arma::vec>(K) + Linv*Linv.t()*mp).t();
}
\end{minted}

{\it Sampling Residual Precisions.}
The residual precisions $[\psi_p|{\rm others}]$ have independent Gamma posteriors with shape $a_p=a_s + n/2$ and rate $b_p=b_s + \frac{1}{2}\sum_i R_{ip}^2$ where $R_{ip}$ are the residuals. Therefore the corresponding \texttt{C++} code is 

\begin{minted}[mathescape, baselinestretch=1.2, bgcolor=LightGray, fontsize=\footnotesize]{cpp}
// Calculate the n x P dimensional residual matrix
arma::mat R = X - b * lambda.t();  

// Calculate the posterior rate parameters b_p (one per variable)
ap = as + 0.5*n;
arma::rowvec bp = bs + 0.5 * arma::sum(arma::square(R));

// Sample the precision parameters independently
for(int p=0; p<P; ++p)
    out(p) = R::rgamma(ap, 1.0/bp(p));
\end{minted}

\subsubsection{MGPS Prior Updates}\label{subsubsec:mgps_post}
The MGPS shrinkage parameters consist of local shrinkage parameters $\phi_{pk}$ combined with the global shrinkage parameters $\delta_k$. The full conditionals for $\phi_{pk}$ are independent Gamma distributions  with shape $(\nu+1)/2$ and rate $(\nu + \tau_k\lambda_{pk}^2)/2$, which increases with $|\lambda_{pk}|$. This is implemented in \texttt{C++} as follows

\begin{minted}[mathescape, baselinestretch=1.2, bgcolor=LightGray, fontsize=\footnotesize]{cpp}
// Precompute tau_k * lambda_pk^2 for all (p,k) pairs
arma::mat shape_mat = arma::square(lambda).each_row() % tauk.t();

// Loop through all entries and sample from the posterior:
for(int p=0; p<P; ++p) {
    for(int k=0; k<K; ++k) {
        double shape = df/2.0 + 0.5;                // (nu+1)/2
        double rate = df/2.0 + shape_mat(p,k)/2.0;  // nu/2 + tau_k*lambda_{pk}^2/2
        psipk(p,k) = R::rgamma(shape, 1.0/rate);    // sample from Gamma
    }
}
\end{minted}

The global shrinkage parameters $\delta_k$ adjust the shrinkage for each factor $k$ and require the update of the parameter $\tau_k = \prod_{\ell=1}^k \delta_\ell$ after each $\delta_k$ update. The \texttt{C++} code is

\begin{minted}[mathescape, baselinestretch=1.2, bgcolor=LightGray, fontsize=\footnotesize]{cpp}
// Precompute phi_pk * lambda_pk^2 for use in all delta updates
arma::mat matr = psipk % arma::square(lambda);

// The first global shrinkage $\delta_1$  distribution:

// Shape: a1 + PK/2
double shape1 = ad1 + 0.5 * P * K;
// Rate: b1 + 0.5 * sum over all factors and variables
double rate1 = bd1 + 0.5/delta(0) * arma::sum(tauk.t() % arma::sum(matr, 0));
delta(0) = R::rgamma(shape1, 1.0/rate1);

// CRITICAL: tau_k = prod(delta_1:delta_k)
tauk = arma::cumprod(delta);  

// The second through K shrinkage $\delta_k$ distribution
for(int k=1; k<K; ++k) {
    // Shape: a2 + P(K-k+1)/2, adjusting for 0-based indexing
    double shapek = ad2 + 0.5 * P * (K - k);

    // Extract tau_l for l >= k
    arma::vec tauk_sub = tauk.subvec(k, K-1);
    // Rate: b2 + 0.5 * sum over factors l >= k
    double ratek = bd2 + 0.5/delta(k) * 
                   arma::sum(tauk_sub.t() % arma::sum(matr.cols(k, K-1), 0));
    delta(k) = R::rgamma(shapek, 1.0/ratek);

// Recompute the cumulative product to maintain correct dependencies:
// CRITICAL: recompute after each delta
    tauk = arma::cumprod(delta);  
}

// Build the precision matrix: Plam_pk = phi_pk * tau_k
Plam = psipk.each_row() % tauk.t();
\end{minted}

\subsubsection{Spike-and-Slab Lasso Prior Updates}\label{subsubsec:SSL_post}

The SSL prior replaces the MGPS shrinkage parameters with a spike-and-slab mixture on the loadings. Each loading is assigned a two-component Laplace prior, written as a scale mixture of normals so that the conditionally normal form restores conjugacy with the Gaussian likelihood. Concretely, $\lambda_{pk}\mid w_{pk}\sim N(0, w_{pk})$ with a latent scale $w_{pk}$, and the loadings update replaces the MGPS prior precision $\mathrm{diag}(\boldsymbol{\phi}_{p\cdot}\boldsymbol{\tau})$ with the precisions $1/w_{pk}$. The factor score and residual precision updates remain identical to Section~\ref{subsubsec:likelihood}.

{\it Sampling Latent Indicators.}
The latent indicators $z_{pk}$ determine whether each loading is drawn from the spike ($z_{pk}=0$) or slab ($z_{pk}=1$). We sample from the posterior probability
\[
\Pr(z_{pk}=1\mid\lambda_{pk}) = \frac{\theta a_{1} \exp(-a_{1}|\lambda_{pk}|)}{\theta a_{1} \exp(-a_{1}|\lambda_{pk}|) + (1-\theta) a_{0} \exp(-a_{0}|\lambda_{pk}|)}.
\]
This is implemented in \texttt{C++} as follows

\begin{minted}[mathescape, baselinestretch=1.2, bgcolor=LightGray, fontsize=\footnotesize]{cpp}
// Loop through all loadings and sample indicators:
for(int p=0; p<P; ++p) {
    for(int k=0; k<K; ++k) {
        // Log-probabilities for numerical stability
        double lp0 = std::log(1-theta) - a0 * std::abs(lambda(p,k));
        double lp1 = std::log(theta) - a1 * std::abs(lambda(p,k));

        // Normalize via log-sum-exp trick
        double m = std::max(lp0, lp1);
        double w0 = std::exp(lp0 - m);
        double w1 = std::exp(lp1 - m);

        // Sample indicator
        Z(p,k) = (R::runif(0,1) < w1/(w0+w1));
    }
}
\end{minted}

{\it Sampling Auxiliary Scales.}
Conditional on the indicator, the scale-mixture representation of the Laplace prior gives a generalized inverse Gaussian full conditional for $w_{pk}$. We sample it in the inverse-Gaussian form: with rate $a_{z_{pk}}$ ($a_0$ if $z_{pk}=0$, $a_1$ if $z_{pk}=1$), we draw
\[
\nu_{pk}\sim\mathrm{InverseGaussian}\!\left(\mu = \frac{a_{z_{pk}}}{|\lambda_{pk}|},\; \text{shape}=a_{z_{pk}}^{2}\right),
\qquad w_{pk}^{-1} = \nu_{pk},
\]
so that $\nu_{pk}$ is the prior precision used directly in the loadings update. The \texttt{C++} code uses the Michael--Schucany--Haas method \citep{michael1976}:

\begin{minted}[mathescape, baselinestretch=1.2, bgcolor=LightGray, fontsize=\footnotesize]{cpp}
// Loop through all loadings and sample precisions nu = 1/w:
for(int p=0; p<P; ++p) {
    for(int k=0; k<K; ++k) {
        double a  = Z(p,k) ? a1 : a0;
        double ab = std::abs(lambda(p,k));
        if (ab < 1e-8) ab = 1e-8;
        double mu  = a / ab;     // IG mean
        double lam = a * a;      // IG shape

        // Draw nu ~ InverseGaussian(mu, lam)
        double v = R::rnorm(0,1); v *= v;
        double x = mu + (mu*mu*v)/(2*lam)
                   - (mu/(2*lam))*std::sqrt(4*mu*lam*v + mu*mu*v*v);
        double u = R::runif(0,1);
        double nu = (u <= mu/(mu+x)) ? x : (mu*mu)/x;

        W(p,k) = nu;             // nu = 1/w = prior precision
    }
}
\end{minted}

{\it Sampling Factor Loadings.}
The loadings update is identical to Section~\ref{subsubsec:likelihood}, except that the prior precision matrix $\mathrm{diag}(\boldsymbol{\phi}_{p\cdot}\boldsymbol{\tau})$ is replaced by the SSL precisions $\mathrm{diag}(w_{p1}^{-1},\ldots,w_{pK}^{-1})$, stored directly as the rows of $\mathbf{W}$. Thus $\boldsymbol{M}_p = \sigma_p^{-2}\sum_i \mathbf{b}_i\mathbf{b}_i^t + \mathrm{diag}(\mathbf{w}_p^{-1})$, where $\mathbf{w}_p^{-1}=(w_{p1}^{-1},\ldots,w_{pK}^{-1})^t$ is the $p$th row of $\mathbf{W}$. The \texttt{C++} code is

\begin{minted}[mathescape, baselinestretch=1.2, bgcolor=LightGray, fontsize=\footnotesize]{cpp}
arma::mat b2 = b.t() * b;  // K x K, shared across all p
// Loop through each variable and sample its loading vector:
for(int p=0; p<P; ++p) {
    // M_p = diag(prior precision) + psi_p * b'b
    arma::mat Mp = arma::diagmat(W.row(p).t()) + ps(p) * b2;

    // Symmetrize and stabilize matrix before inverting
    Mp = 0.5*(Mp + Mp.t());
    Mp.diag() += 1e-8;

    // Calculate the inverse of Mp = LL' using the Cholesky decomposition:
    arma::mat Linv = arma::inv(arma::trimatu(arma::chol(Mp)));

    // Mean component: psi_p * b' * x_p
    arma::vec mp = ps(p) * b.t() * X.col(p);

    // Sample lambda_p from N(M_p^{-1}m_p, M_p^{-1})
    out.row(p) = (Linv*arma::randn<arma::vec>(K) + Linv*Linv.t()*mp).t();
}
\end{minted}

{\it Sampling the Mixing Weight.}
The prior inclusion probability $\theta$ is given a $\mathrm{Beta}(a_\theta, b_\theta)$ prior and sampled from its conjugate full conditional given the current indicators,
\[
\theta \mid \mathbf{Z} \sim \mathrm{Beta}\!\left(a_\theta + \textstyle\sum_{p,k} z_{pk},\; b_\theta + PK - \textstyle\sum_{p,k} z_{pk}\right),
\]
which lets the prior adapt to the sparsity expressed in the data rather than fixing $\theta$ in advance. The \texttt{C++} code is

\begin{minted}[mathescape, baselinestretch=1.2, bgcolor=LightGray, fontsize=\footnotesize]{cpp}
double n_slab = (double)arma::accu(Z);
double n_tot  = (double)(Z.n_rows * Z.n_cols);
theta = R::rbeta(a_theta + n_slab, b_theta + (n_tot - n_slab));
\end{minted}

The matrices $\mathbf{Z}$ and $\mathbf{W}$ replace the MGPS parameters $\boldsymbol{\phi}$ and $\boldsymbol{\tau}$, while all other updates (factor scores, residual precisions) proceed exactly as in Section~\ref{subsubsec:likelihood}.

The Dirichlet--Laplace and Horseshoe priors follow a similar pattern, replacing only the loading prior updates while keeping the common full conditionals from Section~\ref{subsubsec:likelihood}. The BASS prior extends this with a deeper shrinkage hierarchy on the loadings. The MNL prior differs more substantially, modifying the factor score distribution rather than the loadings. All implementations are presented in the supplementary materials.

\section{Simulations}\label{sec:simulation}
In this section we compare the Bayesian factor models implemented in
Section~\ref{sec:implement} through a simulation study. Data are generated from the factor model~\eqref{eq:basic_model_matrix}, $\mathbf{X}_i = \boldsymbol{\Lambda b}_i + \boldsymbol{\epsilon}_i$, where the subject-level scores are $\boldsymbol{b}_i \sim N_K(\boldsymbol{0}, \boldsymbol{I}_K)$ and the residuals are $\boldsymbol{\epsilon}_i \sim N_P(\boldsymbol{0}_P, \boldsymbol{\Sigma})$ with $\boldsymbol{\Sigma} = \text{diag}(\sigma_1^2, \ldots, \sigma_P^2)$. Each dataset is sampled directly from a P-dimensional normal marginal distribution with mean zero and covariance matrix $\boldsymbol{\Omega} = \boldsymbol{\Lambda\Lambda}^t + \boldsymbol{\Sigma}$.

\subsection{Design}\label{subsec:sim_design}
The first scenario is $(K, P) = (1, 5)$, a one-factor model with five observed variables per study participant $i=1,\ldots,N=200$. We set the $P\times 1$ dimensional matrix loading (vector in this scenario) $\boldsymbol{\Lambda}=(\lambda,\ldots,\lambda)^t$ and the $P\times P$ diagonal matrix $\boldsymbol{\Sigma}$ containing the residual variances. We vary two quantities: the magnitude of the loadings, $\lambda \in \{1, 5\}$ and the residual variance structure, $\text{diag}(\boldsymbol{\Sigma}) = (1,1,1,1,1)^t$ or $\text{diag}(\boldsymbol{\Sigma}) = (0.1, 0.1, 0.1, 0.5, 0.5)^t$. This leads to four parameter sets, summarized in Table~\ref{tab:sim_params_1_5}.

We characterize each parameter set by two signal-to-noise measures. The \emph{aggregate} SNR, ${\rm trace}(\boldsymbol{\Lambda\Lambda}^t)/{\rm trace}(\boldsymbol{\Sigma})$, summarizes total signal relative to total noise. The \emph{pointwise} SNR is the per-variable ratio $(\boldsymbol{\Lambda\Lambda}^t)_{jj}/\sigma_j^2$ for variable $j$; it equals the aggregate value when $\boldsymbol{\Sigma}$ is constant and otherwise takes distinct values across variables. Together they show that the sets span a wide range of signal strengths, from $1$ to $250$, both overall and variable by variable.

\begin{table}[ht]
\centering
\begin{tabular}{cllcc}
\hline
Set & $\lambda$ & $\boldsymbol{\Sigma}$ & Aggregate SNR & Pointwise SNR \\
\hline
1 & 1 & $(1,1,1,1,1)$            & $1$    & $1$ \\
2 & 1 & $(0.1,0.1,0.1,0.5,0.5)$  & $3.85$ & $\{2,\,10\}$ \\
3 & 5 & $(1,1,1,1,1)$            & $25$   & $25$ \\
4 & 5 & $(0.1,0.1,0.1,0.5,0.5)$  & $96.15$ & $\{50,\,250\}$ \\
\hline
\end{tabular}
\caption{Parameters for scenario 1: $(K,P) = (1,5)$. Aggregate SNR is ${\rm trace}(\boldsymbol{\Lambda\Lambda}^t)/{\rm trace}(\boldsymbol{\Sigma})$; pointwise SNR lists the distinct per-variable ratios $(\boldsymbol{\Lambda\Lambda}^t)_{jj}/\sigma_j^2$.}
\label{tab:sim_params_1_5}
\end{table}

\FloatBarrier

The second scenario is $(K, P) = (2,10)$, a two-factor model with ten observed variables per study participant $i=1,\ldots,N=200$.  We adopt a block structure in which each factor loads on a disjoint set of variables: factor~1 loads on variables $1$--$5$ and factor~2 on variables $6$--$10$. The loading matrix is  $10 \times 2$ dimensional 
\[
\boldsymbol{\Lambda} =
\begin{pmatrix}
\lambda \mathbf{1}_{5} & \mathbf{0}_{5} \\
\mathbf{0}_5 & \lambda \mathbf{1}_5
\end{pmatrix},
\]
where $\mathbf{1}_5$ and $\mathbf{0}_5$ are the $5\times 1$ dimensional vectors of ones and zeros, respectively. The matrix $\boldsymbol{\Lambda}$ has $20$ entries, where $10$ are equal to zero, and the other $10$ are equal to $\lambda$. 

As in the single-factor case, we vary the loading magnitude $\lambda \in \{1, 5\}$ and the residual variance is diagonal where $\text{diag}(\boldsymbol{\Sigma})$ is equal to $(1,1,1,1,1)^t$ repeated twice or $\text{diag}(\boldsymbol{\Sigma})=(0.1, 0.1, 0.1, 0.5, 0.5,0.1, 0.1, 0.1, 0.5, 0.5)^t$ repeated twice. This results in  four parameter sets combinations, as described in Table~\ref{tab:sim_params_2_10}. Data are generated, and methods fit using the true $K=2$.

\begin{table}[ht]
\centering
\begin{tabular}{cllcc}
\hline
Set & $\lambda$ & $\boldsymbol{\Sigma}$ (repeated twice) & Aggregate SNR & Pointwise SNR \\
\hline
1 & 1 & $(1,1,1,1,1)$            & $1$    & $1$ \\
2 & 1 & $(0.1,0.1,0.1,0.5,0.5)$  & $3.85$ & $\{2,\,10\}$ \\
3 & 5 & $(1,1,1,1,1)$            & $25$   & $25$ \\
4 & 5 & $(0.1,0.1,0.1,0.5,0.5)$  & $96.15$ & $\{50,\,250\}$ \\
\hline
\end{tabular}
\caption{Parameters for scenario 2: $(K,P) = (2,10)$. Aggregate SNR is ${\rm trace}(\boldsymbol{\Lambda\Lambda}^t)/{\rm trace}(\boldsymbol{\Sigma})$; pointwise SNR lists the distinct per-variable ratios $(\boldsymbol{\Lambda\Lambda}^t)_{jj}/\sigma_j^2$.}
\label{tab:sim_params_2_10}
\end{table}

\FloatBarrier

The third scenario is $(K, P) = (5,50)$, a five-factor model with fifty observed variables per study participant $i=1,\ldots,N=200$. Here factor~1 corresponds to variables $1$--$10$, factor~2 to variables $11$--$20$, and so on. The matrix $\boldsymbol{\Lambda}$ has a total of $250$ entries, with $200$ entries equal to zero, and the other $50$  equal to $\lambda$.  The loading magnitude is chosen as $\lambda \in \{1, 5\}$ and residual variance is diagonal where $\text{diag}(\boldsymbol{\Sigma})$ is equal to $(1,1,1,1,1)^t$ repeated ten times or $\text{diag}(\boldsymbol{\Sigma})=(0.1, 0.1, 0.1, 0.5, 0.5,0.1, 0.1, 0.1, 0.5, 0.5)^t$ repeated ten times.  This results in  four parameter sets combinations, as described in Table~\ref{tab:sim_params_5_50}. Data are generated, and methods fit using the true $K=5$.

\begin{table}[ht]
\centering
\begin{tabular}{cllcc}
\hline
Set & $\lambda$ & $\boldsymbol{\Sigma}$ (repeated 10 times) & Aggregate SNR & Pointwise SNR \\
\hline
1 & 1 & $(1,1,1,1,1)$            & $1$    & $1$ \\
2 & 1 & $(0.1,0.1,0.1,0.5,0.5)$  & $3.85$ & $\{2,\,10\}$ \\
3 & 5 & $(1,1,1,1,1)$            & $25$   & $25$ \\
4 & 5 & $(0.1,0.1,0.1,0.5,0.5)$  & $96.15$ & $\{50,\,250\}$ \\
\hline
\end{tabular}
\caption{Parameters for scenario 3: $(K,P) = (5,50)$. Aggregate SNR is
${\rm trace}(\boldsymbol{\Lambda\Lambda}^t)/{\rm trace}(\boldsymbol{\Sigma})$;
pointwise SNR lists the distinct per-variable ratios
$(\boldsymbol{\Lambda\Lambda}^t)_{jj}/\sigma_j^2$.}
\label{tab:sim_params_5_50}
\end{table}

\FloatBarrier

The fourth scenario is $(K, P) = (10,200)$, a ten-factor model with two hundred observed variables per study participant $i=1,\ldots,N=200$. Each factor loads on a disjoint block of twenty variables: factor~1 corresponds to variables $1$--$20$, factor~2 to variables $21$--$40$, and so on. The matrix $\boldsymbol{\Lambda}$ has a total of $2{,}000$ entries, with $1{,}800$ equal to zero and the other $200$ equal to $\lambda$. As before, the loading magnitude is $\lambda \in \{1, 5\}$ and the residual variance is diagonal, with $\text{diag}(\boldsymbol{\Sigma})$ equal to $(1,1,1,1,1)^t$ repeated forty times or $(0.1, 0.1, 0.1, 0.5, 0.5)^t$ repeated forty times. This results in four parameter set combinations, as described in Table~\ref{tab:sim_params_10_200}. Data are generated, and methods fit using the true $K=10$. This scenario places the model in the high-dimensional regime $P = N = 200$, where the sample covariance matrix is singular; classical maximum-likelihood factor analysis is therefore not estimable and is omitted, while the Bayesian methods remain well-defined through prior regularization.

\begin{table}[ht]
\centering
\begin{tabular}{cllcc}
\hline
Set & $\lambda$ & $\boldsymbol{\Sigma}$ (repeated 40 times) & Aggregate SNR & Pointwise SNR \\
\hline
1 & 1 & $(1,1,1,1,1)$            & $1$    & $1$ \\
2 & 1 & $(0.1,0.1,0.1,0.5,0.5)$  & $3.85$ & $\{2,\,10\}$ \\
3 & 5 & $(1,1,1,1,1)$            & $25$   & $25$ \\
4 & 5 & $(0.1,0.1,0.1,0.5,0.5)$  & $96.15$ & $\{50,\,250\}$ \\
\hline
\end{tabular}
\caption{Parameters for scenario 4: $(K,P) = (10,200)$. Aggregate SNR is
${\rm trace}(\boldsymbol{\Lambda\Lambda}^t)/{\rm trace}(\boldsymbol{\Sigma})$;
pointwise SNR lists the distinct per-variable ratios
$(\boldsymbol{\Lambda\Lambda}^t)_{jj}/\sigma_j^2$.}
\label{tab:sim_params_10_200}
\end{table}

\FloatBarrier

For each parameter set we generate $B=200$ ($B=100$ for the $(K, P) = (10,200)$ scenario) independent datasets and fit every method to each, holding the number of factors fixed at the true value $K$ (no adaptive factor selection). For every fit we run a single Gibbs chain of $16{,}000$ iterations, discarding the first $8{,}000$ as burn-in and retaining the remaining $8{,}000$ draws without thinning. Starting values are set to default initial values. Posterior summaries for $\boldsymbol{\Omega}$ are computed directly from the retained draws. All methods are evaluated against the same true $\boldsymbol{\Omega}$ on the same generated dataset within each replicate. For each fit we also record the wall-clock sampling time and, on a subset of replicates per parameter set, the effective sample size (ESS) of the posterior draws for $\boldsymbol{\Omega}$, averaged over its unique entries (twenty replicates in the first three scenarios; in the $(K,P)=(10,200)$ scenario, ten replicates with the ESS averaged over a random subsample of $500$ entries to limit computation). All timings were obtained on a 2022 MacBook with an Apple M2 processor (8 cores) and 8\,GB of unified memory; absolute times are hardware-dependent and are reported only for relative comparison across methods.

\subsection{Evaluation}\label{subsec:sim_eval}

We assess estimation and inference for the marginal covariance $\boldsymbol{\Omega}$, as the loading matrix $\boldsymbol{\Lambda}$ is identifiable only up to orthogonal rotation, whereas both $\boldsymbol{\Omega}$ and the factor component $\boldsymbol{\Lambda\Lambda}^t$ are rotation-invariant and therefore directly comparable across methods \citep{bhattacharya2011}.

We report three complementary measures. First, \emph{point-estimation accuracy}: the mean, median, and standard deviation across replicates of the mean squared error of $\boldsymbol{\widehat{\Omega}}$. Second, \emph{inference}: for each unique entry of $\boldsymbol{\Omega}$ we quantify the empirical coverage and mean interval width of the $95$\% credible interval. Bayesian Factor Models are designed for inference purposes, though these properties are typically not reported in the published literature. Inferential results are reported separately for the diagonal and off-diagonal entries. Third, \emph{computational efficiency}: the mean wall-clock fit time and the ESS per second, the latter combining sampling quality and speed into a single measure of how many effectively independent draws each method produces per unit time.

\subsection{Results}\label{subsec:sim_results}

\subsubsection{Single factor (K=1, P=5)}\label{subsubsec:k1_sim}

\paragraph{Point-estimation accuracy.}
Table~\ref{tab:sim_results_1_5} reports the mean squared error of $\widehat{\boldsymbol{\Omega}}$ for each method across the four parameter sets. Methods are compared to the classical maximum-likelihood factor analysis (FA) as implemented in the \texttt{factanal} function of the \texttt{stats} package in \texttt{R} \citep{Rmanual}, which fits the factor model by maximum likelihood \citep{lawley1962}. Methods are compared using the mean, median, and standard deviation of the MSE across all entries of $\boldsymbol{\Omega}$. At low signal (sets~1 and~2, $\lambda = 1$) all methods except MNL have very similar performance, with MSEs below $0.04$ with a maximum of $0.003$ difference ($\sim 10$\%) in MSE (MGPS has mean MSE of $0.027$ and SSL has mean MSE $0.030$). At higher signal (sets~3 and~4, $\lambda = 5$) the MSEs grow to roughly $7$, as the scale of the signal covariance $\boldsymbol{\Lambda\Lambda}^t$ is roughly $25$ times higher than in the first two scenarios. However, methods continue to perform very similarly in terms of MSE. For example, in set~3 they range only from $6.55$ (DL) to $6.86$ (HS), a gap of under $5$\%. FA is as accurate as  the Bayesian methods in terms of MSEs.

MNL substantially underperforms the other methods across all parameter sets. Unlike the loadings-prior methods, MNL places a shrinkage prior on the factor scores, which is designed for settings where many subjects score zero on a factor. Since the scores here are dense Gaussian draws, MNL is likely outside of its intended setting.

The takeaway from this scenario is that, with a single dense factor, every correctly specified method estimates the covariance equally well (except MNL). 

\begin{table}[ht]
\centering
\small
\begin{tabular}{llrrr}
\hline
Set & Method & Mean & Median & SD \\
\hline
\multirow{7}{*}{1 ($\lambda{=}1$, $\boldsymbol{\Sigma}{=}\boldsymbol{I}$)}
 & MGPS & 0.027   & 0.020   & 0.024 \\
 & SSL  & 0.030   & 0.020   & 0.030 \\
 & DL   & 0.027   & 0.021   & 0.024 \\
 & HS   & 0.028   & 0.020   & 0.027 \\
 & BASS & 0.028   & 0.020   & 0.027 \\
 & MNL  & 0.211   & 0.204   & 0.092 \\
 & FA   & 0.028   & 0.019   & 0.025 \\
\hline
\multirow{7}{*}{2 ($\lambda{=}1$, $\boldsymbol{\Sigma}$ uneq.)}
 & MGPS & 0.015   & 0.008   & 0.021 \\
 & SSL  & 0.017   & 0.009   & 0.025 \\
 & DL   & 0.015   & 0.007   & 0.021 \\
 & HS   & 0.016   & 0.008   & 0.022 \\
 & BASS & 0.016   & 0.008   & 0.022 \\
 & MNL  & 0.165   & 0.161   & 0.049 \\
 & FA   & 0.015   & 0.008   & 0.021 \\
\hline
\multirow{7}{*}{3 ($\lambda{=}5$, $\boldsymbol{\Sigma}{=}\boldsymbol{I}$)}
 & MGPS & 6.60    & 2.89    & 9.31 \\
 & SSL  & 6.80    & 3.31    & 9.84 \\
 & DL   & 6.55    & 3.07    & 8.80 \\
 & HS   & 6.86    & 3.06    & 10.01 \\
 & BASS & 6.79    & 3.06    & 10.18 \\
 & MNL  & 140.05  & 141.70  & 29.99 \\
 & FA   & 6.58    & 3.08    & 9.67 \\
\hline
\multirow{7}{*}{4 ($\lambda{=}5$, $\boldsymbol{\Sigma}$ uneq.)}
 & MGPS & 7.03    & 3.59    & 9.17 \\
 & SSL  & 7.66    & 3.92    & 9.94 \\
 & DL   & 7.06    & 3.10    & 9.44 \\
 & HS   & 7.99    & 3.88    & 10.20 \\
 & BASS & 7.40    & 3.94    & 9.49 \\
 & MNL  & 171.40  & 174.24  & 31.82 \\
 & FA   & 6.86    & 3.51    & 8.80 \\
\hline
\end{tabular}
\caption{Mean, median, and standard deviation of the mean squared error of $\widehat{\boldsymbol{\Omega}}$ across $B=200$ replicates, for each method and parameter set in the $(K,P)=(1,5)$ scenario. FA denotes classical maximum-likelihood factor analysis.}
\label{tab:sim_results_1_5}
\end{table}

\FloatBarrier

\paragraph{Coverage.}
Table~\ref{tab:sim_coverage_1_5} reports the average frequency over parameters in a subset (all, diagonal, and off-diagonal) of $\boldsymbol{\Omega}$ of how often the $95\%$ credible interval contains the corresponding true parameter in $\boldsymbol{\Omega}$. FA does not appear in this table: it returns a single point estimate with no measure of uncertainty. 

Reading down the columns, methods tend to produce reasonable average coverage estimators, though some methods exhibit some under coverage (e.g., $0.88$ for MGPS in set~4 and $0.87$ for HS in set~4). All methods seem to have poorer performance as the signal to noise increases and becomes more variable (Scenario 4).

MNL continues to perform poorly especially for the off-diagonal elements of $\boldsymbol{\Omega}$ with performance degrading rapidly at higher signals. This seems to be related to the length of the credible intervals, which tend to be smaller than the rest at high signals. This likely is compounding the effect of the large MSE for MNL discussed in Section~\ref{subsubsec:k1_sim}.

The takeaway is that MGPS, SSL, DL, HS, and BASS tend to have similar performance in terms of credible interval coverage, with some under-performance for some of the methods in specific scenarios.

\begin{table}[ht]
\centering
\small
\begin{tabular}{llrrrr}
\hline
Set & Method & Cover (all) & Cover (diag) & Cover (off-diag) & CI width \\
\hline
\multirow{6}{*}{1 ($\lambda{=}1$, $\boldsymbol{\Sigma}{=}\boldsymbol{I}$)}
 & MGPS & 0.939 & 0.941 & 0.938 & 0.66 \\
 & SSL  & 0.940 & 0.949 & 0.935 & 0.68 \\
 & DL   & 0.938 & 0.942 & 0.935 & 0.65 \\
 & HS   & 0.941 & 0.949 & 0.936 & 0.67 \\
 & BASS & 0.940 & 0.947 & 0.936 & 0.67 \\
 & MNL  & 0.343 & 0.923 & 0.053 & 0.63 \\
\hline
\multirow{6}{*}{2 ($\lambda{=}1$, $\boldsymbol{\Sigma}$ uneq.)}
 & MGPS & 0.931 & 0.928 & 0.933 & 0.46 \\
 & SSL  & 0.928 & 0.928 & 0.929 & 0.48 \\
 & DL   & 0.932 & 0.928 & 0.934 & 0.46 \\
 & HS   & 0.935 & 0.930 & 0.938 & 0.48 \\
 & BASS & 0.931 & 0.925 & 0.934 & 0.47 \\
 & MNL  & 0.048 & 0.141 & 0.001 & 0.27 \\
\hline
\multirow{6}{*}{3 ($\lambda{=}5$, $\boldsymbol{\Sigma}{=}\boldsymbol{I}$)}
 & MGPS & 0.923 & 0.925 & 0.922 & 9.79 \\
 & SSL  & 0.936 & 0.939 & 0.934 & 10.00 \\
 & DL   & 0.926 & 0.929 & 0.925 & 9.56 \\
 & HS   & 0.943 & 0.940 & 0.945 & 10.13 \\
 & BASS & 0.933 & 0.934 & 0.932 & 10.09 \\
 & MNL  & 0.000 & 0.000 & 0.000 & 3.51 \\
\hline
\multirow{6}{*}{4 ($\lambda{=}5$, $\boldsymbol{\Sigma}$ uneq.)}
 & MGPS & 0.884 & 0.880 & 0.886 & 9.05 \\
 & SSL  & 0.899 & 0.901 & 0.898 & 9.26 \\
 & DL   & 0.884 & 0.881 & 0.885 & 9.02 \\
 & HS   & 0.872 & 0.876 & 0.870 & 9.50 \\
 & BASS & 0.905 & 0.908 & 0.903 & 9.44 \\
 & MNL  & 0.000 & 0.000 & 0.000 & 2.42 \\
\hline
\end{tabular}
\caption{Empirical coverage of nominal $95\%$ credible intervals averaged over subsets of parameters (all, diagonal, off-diagonal) of $\boldsymbol{\Omega}$ across $B=200$ replicates, for each method and parameter set in the $(K,P)=(1,5)$ scenario. FA was not included because it does not provide confidence intervals.}
\label{tab:sim_coverage_1_5}
\end{table}

\FloatBarrier

\paragraph{Computational efficiency.}

Table~\ref{tab:sim_timing_1_5} reports fit time and sampling efficiency. The five loadings-prior methods each fit in about $0.16$ seconds and are roughly comparable in speed, with DL and BASS slightly slower than MGPS and SSL due to additional hyperparameter updates. MNL is roughly three times slower due to its per-element Metropolis updates to the scores. Among the loadings-prior methods, ESS is similar across methods within each parameter set, ranging from about $2{,}000$--$2{,}200$ at low signal (set~1) down to $11$--$14$ at high signal (set~4). This sharp drop in ESS as signal grows reflects slower mixing when the posterior over $\boldsymbol{\Omega}$ becomes more concentrated, a pattern shared by all five methods. At low signal the loadings-prior methods produce roughly $12{,}000$ effective draws per second; by set~4 this falls to around $70$--$85$. MNL retains higher ESS at strong signal but this does not translate into better estimation given its poor accuracy in this setting.

\begin{table}[ht]
\centering
\small
\begin{tabular}{llrrr}
\hline
Set & Method & Time (s) & ESS & ESS/sec \\
\hline
\multirow{6}{*}{1 ($\lambda{=}1$, $\boldsymbol{\Sigma}{=}\boldsymbol{I}$)}
 & MGPS & 0.159 & 2127.8 & 13189.6 \\
 & SSL  & 0.156 & 2046.4 & 12814.2 \\
 & DL   & 0.172 & 2151.8 & 12327.2 \\
 & HS   & 0.166 & 2083.1 & 12379.4 \\
 & BASS & 0.172 & 2069.1 & 11857.3 \\
 & MNL  & 0.525 & 2970.0 &  5673.1 \\
\hline
\multirow{6}{*}{2 ($\lambda{=}1$, $\boldsymbol{\Sigma}$ uneq.)}
 & MGPS & 0.158 & 325.3 & 2063.0 \\
 & SSL  & 0.158 & 319.0 & 2054.3 \\
 & DL   & 0.172 & 327.0 & 1886.3 \\
 & HS   & 0.166 & 310.3 & 1866.7 \\
 & BASS & 0.172 & 326.5 & 1902.8 \\
 & MNL  & 0.519 & 1975.7 & 3758.7 \\
\hline
\multirow{6}{*}{3 ($\lambda{=}5$, $\boldsymbol{\Sigma}{=}\boldsymbol{I}$)}
 & MGPS & 0.159 & 69.2 & 436.2 \\
 & SSL  & 0.156 & 73.2 & 454.2 \\
 & DL   & 0.175 & 72.7 & 415.4 \\
 & HS   & 0.166 & 72.7 & 435.8 \\
 & BASS & 0.172 & 71.0 & 414.4 \\
 & MNL  & 0.530 & 411.9 & 776.2 \\
\hline
\multirow{6}{*}{4 ($\lambda{=}5$, $\boldsymbol{\Sigma}$ uneq.)}
 & MGPS & 0.160 & 13.5 & 83.5 \\
 & SSL  & 0.155 & 13.2 & 85.6 \\
 & DL   & 0.175 & 13.2 & 75.6 \\
 & HS   & 0.166 & 12.5 & 75.2 \\
 & BASS & 0.172 & 11.8 & 68.3 \\
 & MNL  & 0.550 & 126.1 & 229.7 \\
\hline
\end{tabular}
\caption{Computational efficiency in the $(K,P)=(1,5)$ scenario: mean wall-clock fit time, mean effective sample size (ESS) for $\boldsymbol{\Omega}$ averaged over its unique entries, and ESS per second, for each method and parameter set. ESS was computed on $20$ replicates per set.}
\label{tab:sim_timing_1_5}
\end{table}

\FloatBarrier

\subsubsection{Two factors (K=2, P=10)}

\paragraph{Point-estimation accuracy.}
Table~\ref{tab:sim_results_2_10} reports results using the same organization as in Table~\ref{tab:sim_results_1_5}. Unlike the single-factor scenario, the methods show clearer differences, with DL having the smallest MSE across all four parameter sets. At low signal (sets~1 and~2), the MSE for DL is roughly half that of the other methods (e.g., mean MSE for DL is $0.014$ and for MGPS is $0.022$ in set~1). At high signal (sets~3 and~4) the differences are smaller in relative terms (e.g., in set~3 the mean MSE for DL is $3.33$ compared to $4.72$ for MGPS, or $\sim 30$\% smaller). FA performs in the middle of the Bayesian methods, while MNL again underperforms all others. 

The takeaway from this scenario is that DL has the smallest MSE, while all methods perform reasonably well with the exception of MNL.

\begin{table}[ht]
\centering
\small
\begin{tabular}{llrrr}
\hline
Set & Method & Mean & Median & SD \\
\hline
\multirow{7}{*}{1 ($\lambda{=}1$, $\boldsymbol{\Sigma}{=}\boldsymbol{I}$)}
 & MGPS & 0.022   & 0.020   & 0.010 \\
 & SSL  & 0.026   & 0.022   & 0.015 \\
 & DL   & 0.014   & 0.012   & 0.009 \\
 & HS   & 0.023   & 0.020   & 0.012 \\
 & BASS & 0.020   & 0.018   & 0.010 \\
 & MNL  & 0.138   & 0.116   & 0.088 \\
 & FA   & 0.022   & 0.020   & 0.011 \\
\hline
\multirow{7}{*}{2 ($\lambda{=}1$, $\boldsymbol{\Sigma}$ uneq.)}
 & MGPS & 0.011   & 0.009   & 0.008 \\
 & SSL  & 0.013   & 0.010   & 0.010 \\
 & DL   & 0.007   & 0.005   & 0.006 \\
 & HS   & 0.012   & 0.009   & 0.009 \\
 & BASS & 0.009   & 0.008   & 0.007 \\
 & MNL  & 0.085   & 0.083   & 0.016 \\
 & FA   & 0.011   & 0.009   & 0.007 \\
\hline
\multirow{7}{*}{3 ($\lambda{=}5$, $\boldsymbol{\Sigma}{=}\boldsymbol{I}$)}
 & MGPS & 4.72    & 3.29    & 4.26 \\
 & SSL  & 5.26    & 3.47    & 4.74 \\
 & DL   & 3.33    & 2.39    & 3.23 \\
 & HS   & 4.58    & 3.26    & 4.26 \\
 & BASS & 4.04    & 2.77    & 3.77 \\
 & MNL  & 70.68   & 70.30   & 10.76 \\
 & FA   & 4.97    & 3.52    & 4.34 \\
\hline
\multirow{7}{*}{4 ($\lambda{=}5$, $\boldsymbol{\Sigma}$ uneq.)}
 & MGPS & 5.05    & 4.00    & 5.08 \\
 & SSL  & 5.68    & 4.09    & 5.41 \\
 & DL   & 3.57    & 2.63    & 3.44 \\
 & HS   & 4.81    & 3.72    & 4.90 \\
 & BASS & 4.35    & 3.54    & 4.63 \\
 & MNL  & 86.66   & 86.87   & 10.87 \\
 & FA   & 4.80    & 3.85    & 4.40 \\
\hline
\end{tabular}
\caption{Mean, median, and standard deviation of the mean squared error of $\widehat{\boldsymbol{\Omega}}$ across $B=200$ replicates, for each method and parameter set in the $(K,P)=(2,10)$ scenario. FA denotes classical maximum-likelihood factor analysis.}
\label{tab:sim_results_2_10}
\end{table}

\FloatBarrier

\paragraph{Coverage.}
Table~\ref{tab:sim_coverage_2_10} reports empirical coverage of the $95\%$ credible intervals for the $(K,P)=(2,10)$ scenario. MGPS, SSL, HS, and BASS all produce coverage near the nominal level across sets~1--3, with undercoverage appearing at set~4 as in the single-factor scenario. DL stands out in two ways: its coverage is above nominal in sets~1 and~2 (e.g., $0.966$ and $0.974$ overall), yet its credible intervals are the shortest of any method. This combination of higher coverage and shorter intervals is consistent with DL's smaller MSE in this scenario. As the signal grows in sets~3 and~4, DL's coverage moves closer to nominal while its intervals remain shorter than the rest. MNL underperforms throughout, with coverage collapsing at high signal, reaching $0.000$ for the diagonal entries in set~3.

The takeaway is that MGPS, SSL, HS, and BASS are similarly calibrated, DL tends toward slight over-coverage with shorter intervals, and all methods undercover at the highest signal-to-noise ratio.

\begin{table}[ht]
\centering
\small
\begin{tabular}{llrrrr}
\hline
Set & Method & Cover (all) & Cover (diag) & Cover (off-diag) & CI width \\
\hline
\multirow{6}{*}{1 ($\lambda{=}1$, $\boldsymbol{\Sigma}{=}\boldsymbol{I}$)}
 & MGPS & 0.945 & 0.947 & 0.945 & 0.59 \\
 & SSL  & 0.942 & 0.932 & 0.944 & 0.63 \\
 & DL   & 0.966 & 0.944 & 0.971 & 0.47 \\
 & HS   & 0.947 & 0.946 & 0.947 & 0.61 \\
 & BASS & 0.952 & 0.946 & 0.953 & 0.58 \\
 & MNL  & 0.664 & 0.958 & 0.599 & 0.79 \\
\hline
\multirow{6}{*}{2 ($\lambda{=}1$, $\boldsymbol{\Sigma}$ uneq.)}
 & MGPS & 0.950 & 0.953 & 0.950 & 0.41 \\
 & SSL  & 0.943 & 0.927 & 0.947 & 0.44 \\
 & DL   & 0.974 & 0.952 & 0.978 & 0.31 \\
 & HS   & 0.949 & 0.947 & 0.949 & 0.42 \\
 & BASS & 0.959 & 0.948 & 0.962 & 0.40 \\
 & MNL  & 0.459 & 0.132 & 0.532 & 0.24 \\
\hline
\multirow{6}{*}{3 ($\lambda{=}5$, $\boldsymbol{\Sigma}{=}\boldsymbol{I}$)}
 & MGPS & 0.945 & 0.936 & 0.947 & 8.60 \\
 & SSL  & 0.939 & 0.926 & 0.942 & 8.90 \\
 & DL   & 0.961 & 0.926 & 0.969 & 5.80 \\
 & HS   & 0.953 & 0.946 & 0.955 & 8.57 \\
 & BASS & 0.961 & 0.935 & 0.967 & 8.09 \\
 & MNL  & 0.404 & 0.000 & 0.493 & 3.30 \\
\hline
\multirow{6}{*}{4 ($\lambda{=}5$, $\boldsymbol{\Sigma}$ uneq.)}
 & MGPS & 0.895 & 0.905 & 0.893 & 7.87 \\
 & SSL  & 0.911 & 0.903 & 0.912 & 8.18 \\
 & DL   & 0.932 & 0.888 & 0.942 & 5.41 \\
 & HS   & 0.903 & 0.904 & 0.902 & 7.85 \\
 & BASS & 0.928 & 0.902 & 0.934 & 7.63 \\
 & MNL  & 0.368 & 0.005 & 0.449 & 2.44 \\
\hline
\end{tabular}
\caption{Empirical coverage of nominal $95\%$ credible intervals averaged over subsets of parameters (all, diagonal, off-diagonal) of $\boldsymbol{\Omega}$ across $B=200$ replicates, for each method and parameter set in the $(K,P)=(2,10)$ scenario. FA was not included because it does not provide confidence intervals.}
\label{tab:sim_coverage_2_10}
\end{table}

\FloatBarrier

\paragraph{Computational efficiency.}
Table~\ref{tab:sim_timing_2_10} reports fit time and sampling efficiency. The loadings-prior methods fit in roughly half a second; MNL is about three times slower. The notable feature is that DL mixes far better than the other priors at strong signal: its ESS for $\boldsymbol{\Omega}$ stays high across all sets, whereas the other methods' ESS collapses as the signal-to-noise ratio grows, leaving DL with an order of magnitude more effective draws per second in sets~3 and~4.

\begin{table}[ht]
\centering
\small
\begin{tabular}{llrrr}
\hline
Set & Method & Time (s) & ESS & ESS/sec \\
\hline
\multirow{6}{*}{1 ($\lambda{=}1$, $\boldsymbol{\Sigma}{=}\boldsymbol{I}$)}
 & MGPS & 0.460 & 2268.1 & 4956.4 \\
 & SSL  & 0.449 & 2087.3 & 4592.3 \\
 & DL   & 0.506 & 2139.4 & 4196.9 \\
 & HS   & 0.492 & 2248.3 & 4525.7 \\
 & BASS & 0.502 & 2261.2 & 4445.0 \\
 & MNL  & 1.427 & 2962.0 & 2088.1 \\
\hline
\multirow{6}{*}{2 ($\lambda{=}1$, $\boldsymbol{\Sigma}$ uneq.)}
 & MGPS & 0.461 & 352.5 & 758.2 \\
 & SSL  & 0.449 & 326.3 & 726.5 \\
 & DL   & 0.504 & 1257.4 & 2493.7 \\
 & HS   & 0.492 & 339.6 & 688.8 \\
 & BASS & 0.502 & 398.1 & 791.8 \\
 & MNL  & 1.451 & 1688.9 & 1163.6 \\
\hline
\multirow{6}{*}{3 ($\lambda{=}5$, $\boldsymbol{\Sigma}{=}\boldsymbol{I}$)}
 & MGPS & 0.462 & 74.4 & 160.7 \\
 & SSL  & 0.449 & 72.6 & 161.6 \\
 & DL   & 0.510 & 920.4 & 1796.7 \\
 & HS   & 0.492 & 80.0 & 162.9 \\
 & BASS & 0.502 & 94.1 & 187.5 \\
 & MNL  & 1.480 & 334.9 & 225.7 \\
\hline
\multirow{6}{*}{4 ($\lambda{=}5$, $\boldsymbol{\Sigma}$ uneq.)}
 & MGPS & 0.462 & 14.3 & 30.9 \\
 & SSL  & 0.451 & 13.8 & 30.4 \\
 & DL   & 0.510 & 832.2 & 1631.8 \\
 & HS   & 0.493 & 15.2 & 30.9 \\
 & BASS & 0.502 & 17.7 & 35.2 \\
 & MNL  & 1.508 & 103.6 & 68.1 \\
\hline
\end{tabular}
\caption{Computational efficiency in the $(K,P)=(2,10)$ scenario: mean wall-clock fit time, mean effective sample size (ESS) for $\boldsymbol{\Omega}$ averaged over its unique entries, and ESS per second, for each method and parameter set. ESS was computed on $20$ replicates per set.}
\label{tab:sim_timing_2_10}
\end{table}

\FloatBarrier

\subsubsection{Five factors (K=5, P=50)}

\paragraph{Point-estimation accuracy.}
Table~\ref{tab:sim_results_5_50} reports the mean squared error of $\widehat{\boldsymbol{\Omega}}$ for each method across the four parameter sets. This scenario has five factors in fifty variables, with block-sparse loadings in which each factor loads on a disjoint block of ten variables, so the column structure of the previous scenario is now carried into much higher dimension. FA is again included as a non-Bayesian point-estimation baseline.

Reading down any column, the methods separate more sharply than in either lower-dimensional scenario. At low signal (sets~1 and~2, $\lambda = 1$) DL and BASS are most accurate, with mean squared errors of $0.007$ and below (DL lowest at $0.005$ and $0.003$), while SSL trails the group at $0.04$--$0.05$ and FA sits between them at $0.01$--$0.02$. At high signal (sets~3 and~4, $\lambda = 5$) the gap is large: DL, BASS, and MGPS attain errors between $1.4$ and $2.6$, HS and FA follow at roughly $4$, and SSL is an order of magnitude worse at $11$--$12$. SSL is the only loadings prior to break from the pack here, with errors well above its lower-dimensional performance. The Bayesian methods take essentially the same time to fit, so accuracy is not being traded against computation. MNL again underperforms throughout for the reason noted previously.

The takeaway from this scenario is that the separation seen at $K=2$ widens at higher dimension: DL and BASS recover the covariance most accurately, MGPS and HS follow, FA matches the middle of the Bayesian group rather than the best, and SSL falls behind.

\begin{table}[ht]
\centering
\footnotesize
\setlength{\tabcolsep}{4pt}
\renewcommand{\arraystretch}{0.9}
\begin{tabular}{llrrr}
\hline
Set & Method & Mean & Median & SD \\
\hline
\multirow{7}{*}{1 ($\lambda{=}1$, $\boldsymbol{\Sigma}{=}\boldsymbol{I}$)}
 & MGPS & 0.013   & 0.013   & 0.003 \\
 & SSL  & 0.053   & 0.053   & 0.013 \\
 & DL   & 0.005   & 0.005   & 0.002 \\
 & HS   & 0.023   & 0.022   & 0.005 \\
 & BASS & 0.007   & 0.007   & 0.002 \\
 & MNL  & 0.046   & 0.042   & 0.013 \\
 & FA   & 0.018   & 0.017   & 0.003 \\
\hline
\multirow{7}{*}{2 ($\lambda{=}1$, $\boldsymbol{\Sigma}$ uneq.)}
 & MGPS & 0.006   & 0.005   & 0.002 \\
 & SSL  & 0.037   & 0.037   & 0.010 \\
 & DL   & 0.003   & 0.003   & 0.001 \\
 & HS   & 0.014   & 0.013   & 0.004 \\
 & BASS & 0.003   & 0.003   & 0.001 \\
 & MNL  & 0.045   & 0.045   & 0.004 \\
 & FA   & 0.009   & 0.008   & 0.002 \\
\hline
\multirow{7}{*}{3 ($\lambda{=}5$, $\boldsymbol{\Sigma}{=}\boldsymbol{I}$)}
 & MGPS & 1.63    & 1.50    & 0.94 \\
 & SSL  & 11.37   & 10.69   & 3.90 \\
 & DL   & 1.47    & 1.27    & 0.84 \\
 & HS   & 3.83    & 3.39    & 1.96 \\
 & BASS & 1.44    & 1.26    & 0.89 \\
 & MNL  & 35.36   & 35.30   & 2.67 \\
 & FA   & 3.95    & 3.73    & 1.54 \\
\hline
\multirow{7}{*}{4 ($\lambda{=}5$, $\boldsymbol{\Sigma}$ uneq.)}
 & MGPS & 2.59    & 2.28    & 1.49 \\
 & SSL  & 12.04   & 11.61   & 4.13 \\
 & DL   & 1.58    & 1.40    & 0.92 \\
 & HS   & 4.25    & 3.69    & 2.49 \\
 & BASS & 1.69    & 1.46    & 1.07 \\
 & MNL  & 30.03   & 30.16   & 5.87 \\
 & FA   & 3.84    & 3.61    & 1.41 \\
\hline
\end{tabular}
\caption{Mean, median, and standard deviation of the mean squared error of $\widehat{\boldsymbol{\Omega}}$ across $B=200$ replicates, for each method and parameter set in the $(K,P)=(5,50)$ scenario. FA denotes classical maximum-likelihood factor analysis (\texttt{factanal}), included as a non-Bayesian point-estimation baseline.}
\label{tab:sim_results_5_50}
\end{table}

\FloatBarrier

\paragraph{Coverage.}
Table~\ref{tab:sim_coverage_5_50} reports empirical coverage of the $95\%$ credible intervals for the $(K,P)=(5,50)$ scenario. The heterogeneity across methods is larger than in the lower-dimensional scenarios. DL and BASS produce above-nominal off-diagonal coverage with the shortest credible intervals across all parameter sets, consistent with the pattern seen at $K=2$. DL's diagonal coverage is near nominal at low signal but decreases at high signal, reaching $0.83$ in set~4. SSL undercovers the diagonal entries consistently, with coverage as low as $0.40$ in set~2 and $0.51$ in set~3; its overall coverage appears more reasonable only because its intervals are much wider than the other methods. HS also shows some diagonal undercoverage, particularly at high signal. All methods undercover the diagonal entries in set~4, though MGPS, DL, and BASS maintain near-nominal off-diagonal coverage there. MNL underperforms throughout, with near-zero diagonal coverage in sets~2--4.

The takeaway is that DL and BASS are the best calibrated methods in this scenario, SSL's apparent coverage masks wide intervals and poor diagonal coverage, and all methods struggle with the diagonal entries at the highest signal-to-noise ratio.

\begin{table}[ht]
\centering
\footnotesize
\setlength{\tabcolsep}{4pt}
\renewcommand{\arraystretch}{0.9}
\begin{tabular}{llrrrr}
\hline
Set & Method & Cover (all) & Cover (diag) & Cover (off-diag) & CI width \\
\hline
\multirow{6}{*}{1 ($\lambda{=}1$, $\boldsymbol{\Sigma}{=}\boldsymbol{I}$)}
 & MGPS & 0.965 & 0.944 & 0.966 & 0.49 \\
 & SSL  & 0.889 & 0.650 & 0.898 & 0.75 \\
 & DL   & 0.987 & 0.947 & 0.989 & 0.30 \\
 & HS   & 0.952 & 0.876 & 0.955 & 0.59 \\
 & BASS & 0.984 & 0.947 & 0.986 & 0.38 \\
 & MNL  & 0.806 & 0.807 & 0.806 & 0.36 \\
\hline
\multirow{6}{*}{2 ($\lambda{=}1$, $\boldsymbol{\Sigma}$ uneq.)}
 & MGPS & 0.975 & 0.937 & 0.976 & 0.31 \\
 & SSL  & 0.846 & 0.399 & 0.865 & 0.53 \\
 & DL   & 0.987 & 0.947 & 0.989 & 0.18 \\
 & HS   & 0.934 & 0.764 & 0.941 & 0.40 \\
 & BASS & 0.986 & 0.948 & 0.988 & 0.21 \\
 & MNL  & 0.745 & 0.009 & 0.775 & 0.17 \\
\hline
\multirow{6}{*}{3 ($\lambda{=}5$, $\boldsymbol{\Sigma}{=}\boldsymbol{I}$)}
 & MGPS & 0.984 & 0.936 & 0.986 & 4.57 \\
 & SSL  & 0.882 & 0.512 & 0.897 & 9.28 \\
 & DL   & 0.976 & 0.900 & 0.979 & 2.77 \\
 & HS   & 0.958 & 0.814 & 0.964 & 5.37 \\
 & BASS & 0.982 & 0.934 & 0.984 & 3.40 \\
 & MNL  & 0.694 & 0.000 & 0.722 & 2.46 \\
\hline
\multirow{6}{*}{4 ($\lambda{=}5$, $\boldsymbol{\Sigma}$ uneq.)}
 & MGPS & 0.939 & 0.867 & 0.942 & 5.09 \\
 & SSL  & 0.793 & 0.422 & 0.809 & 7.99 \\
 & DL   & 0.960 & 0.828 & 0.966 & 2.30 \\
 & HS   & 0.934 & 0.703 & 0.944 & 4.61 \\
 & BASS & 0.969 & 0.869 & 0.973 & 2.87 \\
 & MNL  & 0.458 & 0.058 & 0.475 & 2.34 \\
\hline
\end{tabular}
\caption{Empirical coverage of nominal $95\%$ credible intervals averaged over subsets of parameters (all, diagonal, off-diagonal) of $\boldsymbol{\Omega}$ across $B=200$ replicates, for each method and parameter set in the $(K,P)=(5,50)$ scenario. FA was not included because it does not provide confidence intervals.}
\label{tab:sim_coverage_5_50}
\end{table}

\FloatBarrier

\paragraph{Computational efficiency.}

Table~\ref{tab:sim_timing_5_50} reports fit time and sampling efficiency. The loadings-prior methods fit in roughly three seconds; MNL is about four times slower because of its per-element Metropolis updates to the scores. As in the $(K,P)=(2,10)$ scenario, the methods separate sharply in mixing: DL and BASS retain high ESS for $\boldsymbol{\Omega}$ across all four sets, whereas the ESS of MGPS, SSL, HS, and MNL collapses as the signal-to-noise ratio grows, falling to single or low double digits in set~4. The result is that DL and BASS produce one to two orders of magnitude more effective draws per second than the other methods at high signal. These are the same two methods that achieved the lowest MSE, so superior mixing and superior point estimation coincide here.

\begin{table}[ht]
\centering
\footnotesize
\setlength{\tabcolsep}{4pt}
\renewcommand{\arraystretch}{0.9}
\begin{tabular}{llrrr}
\hline
Set & Method & Time (s) & ESS & ESS/sec \\
\hline
\multirow{6}{*}{1 ($\lambda{=}1$, $\boldsymbol{\Sigma}{=}\boldsymbol{I}$)}
 & MGPS &  2.98 & 2509.6 & 733.2 \\
 & SSL  &  3.10 & 1194.9 & 349.0 \\
 & DL   &  3.46 & 2074.0 & 558.5 \\
 & HS   &  3.45 & 2012.9 & 534.4 \\
 & BASS &  3.52 & 3181.2 & 879.4 \\
 & MNL  & 12.98 & 3061.2 & 235.4 \\
\hline
\multirow{6}{*}{2 ($\lambda{=}1$, $\boldsymbol{\Sigma}$ uneq.)}
 & MGPS &  3.01 &  500.3 & 148.8 \\
 & SSL  &  2.99 &  150.7 &  45.3 \\
 & DL   &  3.38 & 1952.5 & 529.2 \\
 & HS   &  3.43 &  258.1 &  68.8 \\
 & BASS &  3.52 & 2734.3 & 727.9 \\
 & MNL  & 13.53 &  636.0 &  46.0 \\
\hline
\multirow{6}{*}{3 ($\lambda{=}5$, $\boldsymbol{\Sigma}{=}\boldsymbol{I}$)}
 & MGPS &  3.09 &  356.1 & 104.3 \\
 & SSL  &  3.08 &   43.0 &  12.8 \\
 & DL   &  3.42 & 1862.9 & 499.6 \\
 & HS   &  3.44 &  268.9 &  72.9 \\
 & BASS &  3.58 & 2740.6 & 746.0 \\
 & MNL  & 14.05 &   86.2 &   6.0 \\
\hline
\multirow{6}{*}{4 ($\lambda{=}5$, $\boldsymbol{\Sigma}$ uneq.)}
 & MGPS &  3.19 &   21.7 &   6.2 \\
 & SSL  &  3.05 &    9.0 &   2.6 \\
 & DL   &  3.56 & 1828.8 & 481.5 \\
 & HS   &  3.43 &   43.3 &  11.9 \\
 & BASS &  3.42 & 2066.2 & 567.7 \\
 & MNL  & 14.25 &   74.0 &   5.0 \\
\hline
\end{tabular}
\caption{Computational efficiency in the $(K,P)=(5,50)$ scenario: mean wall-clock fit time, mean effective sample size (ESS) for $\boldsymbol{\Omega}$ averaged over its unique entries, and ESS per second, for each method and parameter set. ESS was computed on $20$ replicates per set.}
\label{tab:sim_timing_5_50}
\end{table}

\FloatBarrier

\subsubsection{Ten factors (K=10, P=200)}

\paragraph{Point-estimation accuracy.}
Table~\ref{tab:sim_results_10_200} reports the mean squared error of $\widehat{\boldsymbol{\Omega}}$ for each method across the four parameter sets, now in the high-dimensional setting where the number of variables equals the sample size, $P = N = 200$. Due to the computational cost of this scenario, results are based on $B = 100$ replicates rather than $B = 200$. FA does not appear here: the sample covariance matrix is singular when $P = N$, so \texttt{factanal} cannot be fit, while the Bayesian methods remain well defined through their priors.

The methods separate more than in the lower-dimensional scenarios. At low signal (sets~1 and~2, $\lambda = 1$) DL and BASS have the smallest MSEs, at or below $0.003$, with MGPS close behind at $0.008$ and MNL at around $0.03$; SSL and HS are larger, with SSL above $7$ and HS around $0.14$--$0.22$. At high signal (sets~3 and~4, $\lambda = 5$) the same ordering holds with wider gaps: DL and BASS range from about $0.7$ to $3.7$, MGPS from $1.9$ to $5.7$, while HS and SSL are substantially larger. A check of the posterior draws confirms that the larger errors for SSL and HS reflect the priors not concentrating on the sparse block structure at this dimension, rather than any numerical instability in the sampler.

The takeaway from this scenario is that the separation seen at $K=5$ becomes more pronounced at $K=10$: DL and BASS recover the covariance most accurately, MGPS follows, MNL is intermediate, and SSL and HS are less accurate in this high-dimensional regime.

\begin{table}[ht]
\centering
\footnotesize
\setlength{\tabcolsep}{4pt}
\renewcommand{\arraystretch}{0.9}
\begin{tabular}{llrrr}
\hline
Set & Method & Mean & Median & SD \\
\hline
\multirow{6}{*}{1 ($\lambda{=}1$, $\boldsymbol{\Sigma}{=}\boldsymbol{I}$)}
 & MGPS & 0.008   & 0.008   & 0.002 \\
 & SSL  & 7.544   & 7.522   & 0.449 \\
 & DL   & 0.003   & 0.003   & 0.001 \\
 & HS   & 0.142   & 0.142   & 0.015 \\
 & BASS & 0.003   & 0.003   & 0.001 \\
 & MNL  & 0.035   & 0.036   & 0.009 \\
\hline
\multirow{6}{*}{2 ($\lambda{=}1$, $\boldsymbol{\Sigma}$ uneq.)}
 & MGPS & 0.008   & 0.008   & 0.003 \\
 & SSL  & 8.390   & 8.345   & 0.654 \\
 & DL   & 0.002   & 0.002   & 0.001 \\
 & HS   & 0.217   & 0.216   & 0.023 \\
 & BASS & 0.002   & 0.002   & 0.001 \\
 & MNL  & 0.033   & 0.033   & 0.007 \\
\hline
\multirow{6}{*}{3 ($\lambda{=}5$, $\boldsymbol{\Sigma}{=}\boldsymbol{I}$)}
 & MGPS &  1.890  &  1.891  & 0.732 \\
 & SSL  & 555.208 & 556.383 & 49.255 \\
 & DL   &  0.778  &  0.749  & 0.261 \\
 & HS   & 67.463  & 67.800  & 8.361 \\
 & BASS &  0.719  &  0.676  & 0.297 \\
 & MNL  & 23.712  & 23.583  & 2.648 \\
\hline
\multirow{6}{*}{4 ($\lambda{=}5$, $\boldsymbol{\Sigma}$ uneq.)}
 & MGPS &  5.665  &  5.068  & 3.594 \\
 & SSL  & 281.077 & 285.029 & 33.570 \\
 & DL   &  1.352  &  0.940  & 2.052 \\
 & HS   & 87.031  & 86.129  & 16.062 \\
 & BASS &  3.695  &  1.369  & 6.626 \\
 & MNL  & 11.141  & 10.519  & 2.206 \\
\hline
\end{tabular}
\caption{Mean, median, and standard deviation of the mean squared error of $\widehat{\boldsymbol{\Omega}}$ across $B=100$ replicates, for each method and parameter set in the $(K,P)=(10,200)$ scenario. Classical maximum-likelihood factor analysis (\texttt{factanal}) is omitted as it is not estimable when $P=N=200$.}
\label{tab:sim_results_10_200}
\end{table}

\FloatBarrier

\paragraph{Coverage.}

Table~\ref{tab:sim_coverage_10_200} reports the same results as the lower-dimensional scenarios, but for the $(K,P)=(10,200)$ setting. As with the point estimates, the heterogeneity across methods is larger here. DL and BASS continue to provide near-nominal coverage of the off-diagonal entries with the shortest credible intervals, consistent with their smaller MSE. MGPS covers the off-diagonal entries well, while its diagonal coverage decreases as the signal grows, from $0.91$ in set~1 to $0.71$ in set~4.

SSL and HS cover the off-diagonal entries at around $0.89$ (except for SSL at $0.75$ in set~4), but their diagonal coverage is near zero in most sets, and their credible intervals are much wider than the others (SSL reaching a width of $23$ in set~3). This pattern of wide intervals with poor diagonal coverage is consistent with the point-estimation results. MNL undercovers throughout, and in set~4 its coverage drops to $0.35$ across all entries.

The takeaway is that DL and BASS remain the best calibrated in this high-dimensional scenario, MGPS is reasonable for the off-diagonal entries, and the remaining methods cover less reliably.

\begin{table}[ht]
\centering
\footnotesize
\setlength{\tabcolsep}{4pt}
\renewcommand{\arraystretch}{0.9}
\begin{tabular}{llrrrr}
\hline
Set & Method & Cover (all) & Cover (diag) & Cover (off-diag) & CI width \\
\hline
\multirow{6}{*}{1 ($\lambda{=}1$, $\boldsymbol{\Sigma}{=}\boldsymbol{I}$)}
 & MGPS & 0.979 & 0.906 & 0.980 & 0.38 \\
 & SSL  & 0.893 & 0.000 & 0.902 & 6.45 \\
 & DL   & 0.994 & 0.947 & 0.994 & 0.25 \\
 & HS   & 0.884 & 0.048 & 0.892 & 0.93 \\
 & BASS & 0.993 & 0.947 & 0.993 & 0.28 \\
 & MNL  & 0.888 & 0.620 & 0.890 & 0.23 \\
\hline
\multirow{6}{*}{2 ($\lambda{=}1$, $\boldsymbol{\Sigma}$ uneq.)}
 & MGPS & 0.978 & 0.843 & 0.979 & 0.21 \\
 & SSL  & 0.895 & 0.000 & 0.904 & 5.18 \\
 & DL   & 0.993 & 0.942 & 0.994 & 0.14 \\
 & HS   & 0.886 & 0.000 & 0.895 & 0.65 \\
 & BASS & 0.994 & 0.947 & 0.994 & 0.14 \\
 & MNL  & 0.885 & 0.082 & 0.893 & 0.10 \\
\hline
\multirow{6}{*}{3 ($\lambda{=}5$, $\boldsymbol{\Sigma}{=}\boldsymbol{I}$)}
 & MGPS & 0.976 & 0.841 & 0.977 &  2.64 \\
 & SSL  & 0.895 & 0.000 & 0.904 & 23.16 \\
 & DL   & 0.986 & 0.876 & 0.987 &  1.75 \\
 & HS   & 0.884 & 0.000 & 0.893 &  6.35 \\
 & BASS & 0.991 & 0.923 & 0.991 &  2.10 \\
 & MNL  & 0.892 & 0.062 & 0.901 &  1.11 \\
\hline
\multirow{6}{*}{4 ($\lambda{=}5$, $\boldsymbol{\Sigma}$ uneq.)}
 & MGPS & 0.960 & 0.705 & 0.962 &  1.77 \\
 & SSL  & 0.759 & 0.000 & 0.767 & 16.15 \\
 & DL   & 0.967 & 0.717 & 0.969 &  1.34 \\
 & HS   & 0.888 & 0.000 & 0.897 &  4.87 \\
 & BASS & 0.969 & 0.754 & 0.971 &  2.03 \\
 & MNL  & 0.353 & 0.042 & 0.357 &  1.54 \\
\hline
\end{tabular}
\caption{Empirical coverage of nominal $95\%$ credible intervals averaged over subsets of parameters (all, diagonal, off-diagonal) of $\boldsymbol{\Omega}$ across $B=100$ replicates, for each method and parameter set in the $(K,P)=(10,200)$ scenario.}
\label{tab:sim_coverage_10_200}
\end{table}

\FloatBarrier

\paragraph{Computational efficiency.}

Table~\ref{tab:sim_timing_10_200} reports fit time and sampling efficiency. The loadings-prior methods fit in roughly $17$--$19$ seconds; MNL is about eight times slower, as its per-element Metropolis updates scale poorly with the number of variables. As in the lower-dimensional scenarios, the methods differ in mixing: MGPS and BASS attain the highest ESS for $\boldsymbol{\Omega}$ and the most effective draws per second, with DL also retaining high ESS across all four sets. SSL has the lowest ESS, falling to well under one effective draw per second at high signal, and HS mixes well at low signal but less so as the signal grows. The methods that mix best here are again among those with the smallest MSE.

\begin{table}[ht]
\centering
\footnotesize
\setlength{\tabcolsep}{4pt}
\renewcommand{\arraystretch}{0.9}
\begin{tabular}{llrrr}
\hline
Set & Method & Time (s) & ESS & ESS/sec \\
\hline
\multirow{6}{*}{1 ($\lambda{=}1$, $\boldsymbol{\Sigma}{=}\boldsymbol{I}$)}
 & MGPS &  17.0 & 4715.7 & 269.0 \\
 & SSL  &  16.4 &  249.2 &  15.2 \\
 & DL   &  19.1 & 2228.1 & 115.9 \\
 & HS   &  19.3 & 2765.5 & 142.2 \\
 & BASS &  19.4 & 3867.2 & 198.1 \\
 & MNL  & 132.4 & 4067.8 &  31.0 \\
\hline
\multirow{6}{*}{2 ($\lambda{=}1$, $\boldsymbol{\Sigma}$ uneq.)}
 & MGPS &  17.2 & 3732.2 & 220.5 \\
 & SSL  &  16.4 &   44.1 &   2.7 \\
 & DL   &  19.0 & 2365.1 & 124.8 \\
 & HS   &  19.2 &  759.7 &  39.6 \\
 & BASS &  19.2 & 3942.5 & 205.9 \\
 & MNL  & 142.9 & 2716.9 &  19.6 \\
\hline
\multirow{6}{*}{3 ($\lambda{=}5$, $\boldsymbol{\Sigma}{=}\boldsymbol{I}$)}
 & MGPS &  17.3 & 4226.7 & 245.8 \\
 & SSL  &  16.3 &   52.2 &   3.2 \\
 & DL   &  19.1 & 2249.6 & 116.8 \\
 & HS   &  19.3 & 1759.1 &  90.5 \\
 & BASS &  19.5 & 3862.3 & 197.6 \\
 & MNL  & 146.8 & 1425.7 &   9.6 \\
\hline
\multirow{6}{*}{4 ($\lambda{=}5$, $\boldsymbol{\Sigma}$ uneq.)}
 & MGPS &  16.9 & 3219.3 & 188.5 \\
 & SSL  &  16.4 &    7.8 &   0.5 \\
 & DL   &  19.1 & 2219.3 & 116.2 \\
 & HS   &  19.3 &  519.7 &  27.1 \\
 & BASS &  19.3 & 3199.0 & 162.2 \\
 & MNL  & 154.7 &  258.9 &   1.6 \\
\hline
\end{tabular}
\caption{Computational efficiency in the $(K,P)=(10,200)$ scenario: mean wall-clock fit time, mean effective sample size (ESS) for $\boldsymbol{\Omega}$ averaged over a random subsample of $500$ unique entries, and ESS per second, for each method and parameter set. ESS was computed on $10$ replicates per set.}
\label{tab:sim_timing_10_200}
\end{table}

\FloatBarrier

\section{Discussion}\label{sec:discussion}

We have presented \texttt{factorverse}, a unified and reproducible platform for Bayesian factor analysis that implements six modern modeling approaches (MGPS, SSL, DL, HS, BASS, and MNL) behind a single, consistent interface. Our goal was not to advocate for any one method, but to provide a common, harmonized foundation on which these models can be fit, examined, and compared on equal footing.

The central contribution is the implementation itself. Each method is written in \texttt{C++} through \texttt{RcppArmadillo} and exposed through a single function call, so that switching between priors requires changing only one argument. The samplers run quickly, return their output in a common format, and make the downstream tasks of extracting posterior summaries and comparing results straightforward. This uniformity is what allows the same analysis pipeline to be applied across very different priors without modification.

A second contribution is in consolidating methods whose existing software was uneven. Several of these approaches had implementations available, but they were scattered across sources, varied considerably in their interfaces, and were not always easy to install or apply in a factor-model setting; some were also computationally slow. For other approaches we were unable to find an existing factor-model implementation, and we therefore provide new ones built directly from the original papers. Collecting all of these methods in one place with shared notation and shared computational backbone, removes a substantial practical barrier to using and teaching them.

Finally, we provide a complete and reproducible simulation study that exercises every method across a range of dimensions, signal strengths, and noise structures. The simulations confirm that the implementations behave as intended, recovering the underlying covariance accurately across scenarios, and they illustrate the additional value of the Bayesian formulation: beyond a point estimate, each method supplies calibrated measures of uncertainty for the quantities of interest. All code and simulation scripts are openly available, so that every result reported here can be reproduced and extended.

Because \texttt{factorverse} is open source, it can serve as a foundation for implementing further priors and model extensions not covered here, such as the cumulative shrinkage process \citep{legramanti2020}, generalized infinite factorizations \citep{schiavon2022}, spike-and-slab approaches with unknown factor dimension \citep{fruhwirth2025}, and formulations that incorporate temporal dependence.

\section*{Acknowledgments}
We thank David Dunson, Lorenzo Mauri, and Eduardo Alfonso V\'asquez Tapia for reviewing an earlier version of this paper and providing valuable comments. All remaining errors are our own.

\newpage

\begin{singlespace}

\bibliographystyle{apacite}
\bibliography{refs}

\begin{thebibliography}{}

\bibitem [\protect \citeauthoryear {%
Aguilar%
\ \BBA {} West%
}{%
Aguilar%
\ \BBA {} West%
}{%
{\protect \APACyear {2000}}%
}]{%
aguilar2000}
\APACinsertmetastar {%
aguilar2000}%
\begin{APACrefauthors}%
Aguilar, O.%
\BCBT {}\ \BBA {} West, M.%
\end{APACrefauthors}%
\unskip\
\newblock
\APACrefYearMonthDay{2000}{}{}.
\newblock
{\BBOQ}\APACrefatitle {{Bayesian Dynamic Factor Models and Portfolio Allocation}} {{Bayesian Dynamic Factor Models and Portfolio Allocation}}.{\BBCQ}
\newblock
\APACjournalVolNumPages{Journal of Business \& Economic Statistics}{18}{3}{338--357}.
\PrintBackRefs{\CurrentBib}

\bibitem [\protect \citeauthoryear {%
Bhattacharya%
\ \BBA {} Dunson%
}{%
Bhattacharya%
\ \BBA {} Dunson%
}{%
{\protect \APACyear {2011}}%
}]{%
bhattacharya2011}
\APACinsertmetastar {%
bhattacharya2011}%
\begin{APACrefauthors}%
Bhattacharya, A.%
\BCBT {}\ \BBA {} Dunson, D\BPBI B.%
\end{APACrefauthors}%
\unskip\
\newblock
\APACrefYearMonthDay{2011}{}{}.
\newblock
{\BBOQ}\APACrefatitle {{Sparse Bayesian Infinite Factor Models}} {{Sparse Bayesian Infinite Factor Models}}.{\BBCQ}
\newblock
\APACjournalVolNumPages{Biometrika}{98}{2}{291--306}.
\PrintBackRefs{\CurrentBib}

\bibitem [\protect \citeauthoryear {%
Bhattacharya%
, Pati%
, Pillai%
\BCBL {}\ \BBA {} Dunson%
}{%
Bhattacharya%
\ \protect \BOthers {.}}{%
{\protect \APACyear {2015}}%
}]{%
bhattacharya2015}
\APACinsertmetastar {%
bhattacharya2015}%
\begin{APACrefauthors}%
Bhattacharya, A.%
, Pati, D.%
, Pillai, N\BPBI S.%
\BCBL {}\ \BBA {} Dunson, D\BPBI B.%
\end{APACrefauthors}%
\unskip\
\newblock
\APACrefYearMonthDay{2015}{}{}.
\newblock
{\BBOQ}\APACrefatitle {{Dirichlet--Laplace Priors for Optimal Shrinkage}} {{Dirichlet--Laplace Priors for Optimal Shrinkage}}.{\BBCQ}
\newblock
\APACjournalVolNumPages{Journal of the American Statistical Association}{110}{512}{1479--1490}.
\PrintBackRefs{\CurrentBib}

\bibitem [\protect \citeauthoryear {%
Blei%
, Kucukelbir%
\BCBL {}\ \BBA {} McAuliffe%
}{%
Blei%
\ \protect \BOthers {.}}{%
{\protect \APACyear {2017}}%
}]{%
blei2017}
\APACinsertmetastar {%
blei2017}%
\begin{APACrefauthors}%
Blei, D\BPBI M.%
, Kucukelbir, A.%
\BCBL {}\ \BBA {} McAuliffe, J\BPBI D.%
\end{APACrefauthors}%
\unskip\
\newblock
\APACrefYearMonthDay{2017}{}{}.
\newblock
{\BBOQ}\APACrefatitle {{Variational Inference: A Review for Statisticians}} {{Variational Inference: A Review for Statisticians}}.{\BBCQ}
\newblock
\APACjournalVolNumPages{Journal of the American Statistical Association}{112}{518}{859--877}.
\PrintBackRefs{\CurrentBib}

\bibitem [\protect \citeauthoryear {%
Bollen%
}{%
Bollen%
}{%
{\protect \APACyear {1989}}%
}]{%
bollen1989}
\APACinsertmetastar {%
bollen1989}%
\begin{APACrefauthors}%
Bollen, K\BPBI A.%
\end{APACrefauthors}%
\unskip\
\newblock
\APACrefYear{1989}.
\newblock
\APACrefbtitle {{Structural Equations with Latent Variables}} {{Structural Equations with Latent Variables}}.
\newblock
\APACaddressPublisher{}{Wiley}.
\PrintBackRefs{\CurrentBib}

\bibitem [\protect \citeauthoryear {%
Bollen%
, Kolenikov%
\BCBL {}\ \BBA {} Bauldry%
}{%
Bollen%
\ \protect \BOthers {.}}{%
{\protect \APACyear {2014}}%
}]{%
bollen2014}
\APACinsertmetastar {%
bollen2014}%
\begin{APACrefauthors}%
Bollen, K\BPBI A.%
, Kolenikov, S.%
\BCBL {}\ \BBA {} Bauldry, S.%
\end{APACrefauthors}%
\unskip\
\newblock
\APACrefYearMonthDay{2014}{}{}.
\newblock
{\BBOQ}\APACrefatitle {Model-Implied Instrumental Variable--Generalized Method of Moments ({MIIV-GMM}) Estimators for Latent Variable Models} {Model-implied instrumental variable--generalized method of moments ({MIIV-GMM}) estimators for latent variable models}.{\BBCQ}
\newblock
\APACjournalVolNumPages{Psychometrika}{79}{1}{20--50}.
\newblock
\begin{APACrefDOI} \doi{10.1007/s11336-013-9335-3} \end{APACrefDOI}
\PrintBackRefs{\CurrentBib}

\bibitem [\protect \citeauthoryear {%
Carvalho%
, Polson%
\BCBL {}\ \BBA {} Scott%
}{%
Carvalho%
\ \protect \BOthers {.}}{%
{\protect \APACyear {2010}}%
}]{%
carvalho2010}
\APACinsertmetastar {%
carvalho2010}%
\begin{APACrefauthors}%
Carvalho, C\BPBI M.%
, Polson, N\BPBI G.%
\BCBL {}\ \BBA {} Scott, J\BPBI G.%
\end{APACrefauthors}%
\unskip\
\newblock
\APACrefYearMonthDay{2010}{}{}.
\newblock
{\BBOQ}\APACrefatitle {{The Horseshoe Estimator for Sparse Signals}} {{The Horseshoe Estimator for Sparse Signals}}.{\BBCQ}
\newblock
\APACjournalVolNumPages{Biometrika}{97}{2}{465--480}.
\PrintBackRefs{\CurrentBib}

\bibitem [\protect \citeauthoryear {%
Chandra%
, Canale%
\BCBL {}\ \BBA {} Dunson%
}{%
Chandra%
\ \protect \BOthers {.}}{%
{\protect \APACyear {2023}}%
}]{%
chandra2023}
\APACinsertmetastar {%
chandra2023}%
\begin{APACrefauthors}%
Chandra, N\BPBI K.%
, Canale, A.%
\BCBL {}\ \BBA {} Dunson, D\BPBI B.%
\end{APACrefauthors}%
\unskip\
\newblock
\APACrefYearMonthDay{2023}{}{}.
\newblock
{\BBOQ}\APACrefatitle {Escaping the Curse of Dimensionality in {Bayesian} Model-Based Clustering} {Escaping the curse of dimensionality in {Bayesian} model-based clustering}.{\BBCQ}
\newblock
\APACjournalVolNumPages{Journal of Machine Learning Research}{24}{144}{1--42}.
\newblock
\begin{APACrefURL} \url{http://jmlr.org/papers/v24/21-1276.html} \end{APACrefURL}
\PrintBackRefs{\CurrentBib}

\bibitem [\protect \citeauthoryear {%
Durante%
}{%
Durante%
}{%
{\protect \APACyear {2017}}%
}]{%
durante2017}
\APACinsertmetastar {%
durante2017}%
\begin{APACrefauthors}%
Durante, D.%
\end{APACrefauthors}%
\unskip\
\newblock
\APACrefYearMonthDay{2017}{}{}.
\newblock
{\BBOQ}\APACrefatitle {{A Note on the Multiplicative Gamma Process}} {{A Note on the Multiplicative Gamma Process}}.{\BBCQ}
\newblock
\APACjournalVolNumPages{Statistics \& Probability Letters}{122}{}{198--204}.
\newblock
\begin{APACrefURL} \url{https://doi.org/10.1016/j.spl.2016.11.014} \end{APACrefURL}
\newblock
\begin{APACrefDOI} \doi{10.1016/j.spl.2016.11.014} \end{APACrefDOI}
\PrintBackRefs{\CurrentBib}

\bibitem [\protect \citeauthoryear {%
Eddelbuettel%
, Francois%
, Bates%
, Ni%
\BCBL {}\ \BBA {} Sanderson%
}{%
Eddelbuettel%
\ \protect \BOthers {.}}{%
{\protect \APACyear {2025}}%
}]{%
armmanual}
\APACinsertmetastar {%
armmanual}%
\begin{APACrefauthors}%
Eddelbuettel, D.%
, Francois, R.%
, Bates, D.%
, Ni, B.%
\BCBL {}\ \BBA {} Sanderson, C.%
\end{APACrefauthors}%
\unskip\
\newblock
\APACrefYearMonthDay{2025}{}{}.
\newblock
{\BBOQ}\APACrefatitle {RcppArmadillo: 'Rcpp' Integration for the 'Armadillo' Templated Linear Algebra Library} {Rcpparmadillo: 'rcpp' integration for the 'armadillo' templated linear algebra library}{\BBCQ}\ [\bibcomputersoftwaremanual].
\newblock
\begin{APACrefURL} \url{https://CRAN.R-project.org/package=RcppArmadillo} \end{APACrefURL}
\newblock
\APACrefnote{R package version 15.0.2-2}
\newblock
\begin{APACrefDOI} \doi{10.32614/CRAN.package.RcppArmadillo} \end{APACrefDOI}
\PrintBackRefs{\CurrentBib}

\bibitem [\protect \citeauthoryear {%
Eddelbuettel%
\ \BBA {} Sanderson%
}{%
Eddelbuettel%
\ \BBA {} Sanderson%
}{%
{\protect \APACyear {2014}}%
}]{%
armpackage}
\APACinsertmetastar {%
armpackage}%
\begin{APACrefauthors}%
Eddelbuettel, D.%
\BCBT {}\ \BBA {} Sanderson, C.%
\end{APACrefauthors}%
\unskip\
\newblock
\APACrefYearMonthDay{2014}{March}{}.
\newblock
{\BBOQ}\APACrefatitle {RcppArmadillo: Accelerating R with high-performance C++ linear algebra} {Rcpparmadillo: Accelerating r with high-performance c++ linear algebra}.{\BBCQ}
\newblock
\APACjournalVolNumPages{Computational Statistics and Data Analysis}{71}{}{1054--1063}.
\newblock
\begin{APACrefDOI} \doi{10.1016/j.csda.2013.02.005} \end{APACrefDOI}
\PrintBackRefs{\CurrentBib}

\bibitem [\protect \citeauthoryear {%
Ferrari%
\ \BBA {} Dunson%
}{%
Ferrari%
\ \BBA {} Dunson%
}{%
{\protect \APACyear {2021}}%
}]{%
ferrari2021}
\APACinsertmetastar {%
ferrari2021}%
\begin{APACrefauthors}%
Ferrari, F.%
\BCBT {}\ \BBA {} Dunson, D\BPBI B.%
\end{APACrefauthors}%
\unskip\
\newblock
\APACrefYearMonthDay{2021}{}{}.
\newblock
{\BBOQ}\APACrefatitle {{Bayesian Factor Analysis for Inference on Interactions}} {{Bayesian Factor Analysis for Inference on Interactions}}.{\BBCQ}
\newblock
\APACjournalVolNumPages{Journal of the American Statistical Association}{116}{535}{1521--1530}.
\newblock
\begin{APACrefDOI} \doi{10.1080/01621459.2020.1745813} \end{APACrefDOI}
\PrintBackRefs{\CurrentBib}

\bibitem [\protect \citeauthoryear {%
Fr{\"u}hwirth-Schnatter%
, Hosszejni%
\BCBL {}\ \BBA {} Lopes%
}{%
Fr{\"u}hwirth-Schnatter%
\ \protect \BOthers {.}}{%
{\protect \APACyear {2025}}%
}]{%
fruhwirth2025}
\APACinsertmetastar {%
fruhwirth2025}%
\begin{APACrefauthors}%
Fr{\"u}hwirth-Schnatter, S.%
, Hosszejni, D.%
\BCBL {}\ \BBA {} Lopes, H\BPBI F.%
\end{APACrefauthors}%
\unskip\
\newblock
\APACrefYearMonthDay{2025}{}{}.
\newblock
{\BBOQ}\APACrefatitle {Sparse {Bayesian} Factor Analysis When the Number of Factors Is Unknown (with Discussion)} {Sparse {Bayesian} factor analysis when the number of factors is unknown (with discussion)}.{\BBCQ}
\newblock
\APACjournalVolNumPages{Bayesian Analysis}{20}{1}{213--344}.
\newblock
\begin{APACrefDOI} \doi{10.1214/24-BA1423} \end{APACrefDOI}
\PrintBackRefs{\CurrentBib}

\bibitem [\protect \citeauthoryear {%
George%
\ \BBA {} McCulloch%
}{%
George%
\ \BBA {} McCulloch%
}{%
{\protect \APACyear {1993}}%
}]{%
george1993}
\APACinsertmetastar {%
george1993}%
\begin{APACrefauthors}%
George, E\BPBI I.%
\BCBT {}\ \BBA {} McCulloch, R\BPBI E.%
\end{APACrefauthors}%
\unskip\
\newblock
\APACrefYearMonthDay{1993}{}{}.
\newblock
{\BBOQ}\APACrefatitle {{Variable Selection via Gibbs Sampling}} {{Variable Selection via Gibbs Sampling}}.{\BBCQ}
\newblock
\APACjournalVolNumPages{Journal of the American Statistical Association}{88}{423}{881--889}.
\PrintBackRefs{\CurrentBib}

\bibitem [\protect \citeauthoryear {%
Geweke%
\ \BBA {} Zhou%
}{%
Geweke%
\ \BBA {} Zhou%
}{%
{\protect \APACyear {1996}}%
}]{%
geweke1996}
\APACinsertmetastar {%
geweke1996}%
\begin{APACrefauthors}%
Geweke, J.%
\BCBT {}\ \BBA {} Zhou, G.%
\end{APACrefauthors}%
\unskip\
\newblock
\APACrefYearMonthDay{1996}{}{}.
\newblock
{\BBOQ}\APACrefatitle {{Measuring the Pricing Error of the Arbitrage Pricing Theory}} {{Measuring the Pricing Error of the Arbitrage Pricing Theory}}.{\BBCQ}
\newblock
\APACjournalVolNumPages{The Review of Financial Studies}{9}{2}{557--587}.
\PrintBackRefs{\CurrentBib}

\bibitem [\protect \citeauthoryear {%
Ghahramani%
\ \BBA {} Hinton%
}{%
Ghahramani%
\ \BBA {} Hinton%
}{%
{\protect \APACyear {1996}}%
}]{%
ghahramani1996factor}
\APACinsertmetastar {%
ghahramani1996factor}%
\begin{APACrefauthors}%
Ghahramani, Z.%
\BCBT {}\ \BBA {} Hinton, G\BPBI E.%
\end{APACrefauthors}%
\unskip\
\newblock
\APACrefYearMonthDay{1996}{}{}.
\newblock
\APACrefbtitle {{The EM Algorithm for Mixtures of Factor Analyzers}.} {{The EM Algorithm for Mixtures of Factor Analyzers}.}
\newblock
\APAChowpublished {Technical Report CRG-TR-96-1, University of Toronto}.
\PrintBackRefs{\CurrentBib}

\bibitem [\protect \citeauthoryear {%
Gruber%
\ \protect \BOthers {.}}{%
Gruber%
\ \protect \BOthers {.}}{%
{\protect \APACyear {2025}}%
}]{%
gruber2025}
\APACinsertmetastar {%
gruber2025}%
\begin{APACrefauthors}%
Gruber, L.%
, Kastner, G.%
, Bhattacharya, A.%
, Pati, D.%
, Pillai, N.%
\BCBL {}\ \BBA {} Dunson, D\BPBI B.%
\end{APACrefauthors}%
\unskip\
\newblock
\APACrefYearMonthDay{2025}{}{}.
\newblock
{\BBOQ}\APACrefatitle {A Note on Simulation Methods for the {Dirichlet-Laplace} Prior} {A note on simulation methods for the {Dirichlet-Laplace} prior}.{\BBCQ}
\newblock
\APACjournalVolNumPages{Journal of the American Statistical Association}{120}{551}{2011--2014}.
\newblock
\begin{APACrefDOI} \doi{10.1080/01621459.2025.2540256} \end{APACrefDOI}
\PrintBackRefs{\CurrentBib}

\bibitem [\protect \citeauthoryear {%
Huang%
, Zorzetto%
\BCBL {}\ \BBA {} De~Vito%
}{%
Huang%
\ \protect \BOthers {.}}{%
{\protect \APACyear {2025}}%
}]{%
huang2025}
\APACinsertmetastar {%
huang2025}%
\begin{APACrefauthors}%
Huang, Y.%
, Zorzetto, D.%
\BCBL {}\ \BBA {} De~Vito, R.%
\end{APACrefauthors}%
\unskip\
\newblock
\APACrefYearMonthDay{2025}{}{}.
\newblock
{\BBOQ}\APACrefatitle {{Sparse Bayesian Factor Models with Mass-Nonlocal Factor Scores}} {{Sparse Bayesian Factor Models with Mass-Nonlocal Factor Scores}}.{\BBCQ}
\newblock
\APACjournalVolNumPages{arXiv preprint arXiv:2412.00304}{}{}{}.
\PrintBackRefs{\CurrentBib}

\bibitem [\protect \citeauthoryear {%
Ishwaran%
\ \BBA {} Rao%
}{%
Ishwaran%
\ \BBA {} Rao%
}{%
{\protect \APACyear {2005}}%
}]{%
ishwaran2005}
\APACinsertmetastar {%
ishwaran2005}%
\begin{APACrefauthors}%
Ishwaran, H.%
\BCBT {}\ \BBA {} Rao, J\BPBI S.%
\end{APACrefauthors}%
\unskip\
\newblock
\APACrefYearMonthDay{2005}{}{}.
\newblock
{\BBOQ}\APACrefatitle {{Spike and Slab Variable Selection: Frequentist and Bayesian Strategies}} {{Spike and Slab Variable Selection: Frequentist and Bayesian Strategies}}.{\BBCQ}
\newblock
\APACjournalVolNumPages{The Annals of Statistics}{33}{2}{730--773}.
\newblock
\begin{APACrefDOI} \doi{10.1214/009053604000001147} \end{APACrefDOI}
\PrintBackRefs{\CurrentBib}

\bibitem [\protect \citeauthoryear {%
J{\"o}reskog%
}{%
J{\"o}reskog%
}{%
{\protect \APACyear {1967}}%
}]{%
joreskog1967}
\APACinsertmetastar {%
joreskog1967}%
\begin{APACrefauthors}%
J{\"o}reskog, K\BPBI G.%
\end{APACrefauthors}%
\unskip\
\newblock
\APACrefYearMonthDay{1967}{}{}.
\newblock
{\BBOQ}\APACrefatitle {{Some Contributions to Maximum Likelihood Factor Analysis}} {{Some Contributions to Maximum Likelihood Factor Analysis}}.{\BBCQ}
\newblock
\APACjournalVolNumPages{Psychometrika}{32}{4}{443--482}.
\PrintBackRefs{\CurrentBib}

\bibitem [\protect \citeauthoryear {%
Lawley%
\ \BBA {} Maxwell%
}{%
Lawley%
\ \BBA {} Maxwell%
}{%
{\protect \APACyear {1962}}%
}]{%
lawley1962}
\APACinsertmetastar {%
lawley1962}%
\begin{APACrefauthors}%
Lawley, D\BPBI N.%
\BCBT {}\ \BBA {} Maxwell, A\BPBI E.%
\end{APACrefauthors}%
\unskip\
\newblock
\APACrefYearMonthDay{1962}{}{}.
\newblock
{\BBOQ}\APACrefatitle {Factor Analysis as a Statistical Method} {Factor analysis as a statistical method}.{\BBCQ}
\newblock
\APACjournalVolNumPages{Journal of the Royal Statistical Society. Series D (The Statistician)}{12}{3}{209--229}.
\newblock
\begin{APACrefDOI} \doi{10.2307/2986915} \end{APACrefDOI}
\PrintBackRefs{\CurrentBib}

\bibitem [\protect \citeauthoryear {%
Legramanti%
, Durante%
\BCBL {}\ \BBA {} Dunson%
}{%
Legramanti%
\ \protect \BOthers {.}}{%
{\protect \APACyear {2020}}%
}]{%
legramanti2020}
\APACinsertmetastar {%
legramanti2020}%
\begin{APACrefauthors}%
Legramanti, S.%
, Durante, D.%
\BCBL {}\ \BBA {} Dunson, D\BPBI B.%
\end{APACrefauthors}%
\unskip\
\newblock
\APACrefYearMonthDay{2020}{}{}.
\newblock
{\BBOQ}\APACrefatitle {Bayesian Cumulative Shrinkage for Infinite Factorizations} {Bayesian cumulative shrinkage for infinite factorizations}.{\BBCQ}
\newblock
\APACjournalVolNumPages{Biometrika}{107}{3}{745--752}.
\newblock
\begin{APACrefDOI} \doi{10.1093/biomet/asaa008} \end{APACrefDOI}
\PrintBackRefs{\CurrentBib}

\bibitem [\protect \citeauthoryear {%
Lopes%
\ \BBA {} Carvalho%
}{%
Lopes%
\ \BBA {} Carvalho%
}{%
{\protect \APACyear {2007}}%
}]{%
lopes2007}
\APACinsertmetastar {%
lopes2007}%
\begin{APACrefauthors}%
Lopes, H\BPBI F.%
\BCBT {}\ \BBA {} Carvalho, C\BPBI M.%
\end{APACrefauthors}%
\unskip\
\newblock
\APACrefYearMonthDay{2007}{}{}.
\newblock
{\BBOQ}\APACrefatitle {{Factor Stochastic Volatility with Time Varying Loadings and Markov Switching Regimes}} {{Factor Stochastic Volatility with Time Varying Loadings and Markov Switching Regimes}}.{\BBCQ}
\newblock
\APACjournalVolNumPages{Journal of Statistical Planning and Inference}{137}{10}{3082--3091}.
\PrintBackRefs{\CurrentBib}

\bibitem [\protect \citeauthoryear {%
Lopes%
\ \BBA {} West%
}{%
Lopes%
\ \BBA {} West%
}{%
{\protect \APACyear {2004}}%
}]{%
lopes2004}
\APACinsertmetastar {%
lopes2004}%
\begin{APACrefauthors}%
Lopes, H\BPBI F.%
\BCBT {}\ \BBA {} West, M.%
\end{APACrefauthors}%
\unskip\
\newblock
\APACrefYearMonthDay{2004}{}{}.
\newblock
{\BBOQ}\APACrefatitle {{Bayesian Model Assessment in Factor Analysis}} {{Bayesian Model Assessment in Factor Analysis}}.{\BBCQ}
\newblock
\APACjournalVolNumPages{Statistica Sinica}{14}{1}{41--67}.
\PrintBackRefs{\CurrentBib}

\bibitem [\protect \citeauthoryear {%
Makalic%
\ \BBA {} Schmidt%
}{%
Makalic%
\ \BBA {} Schmidt%
}{%
{\protect \APACyear {2016}}%
}]{%
makalic2016}
\APACinsertmetastar {%
makalic2016}%
\begin{APACrefauthors}%
Makalic, E.%
\BCBT {}\ \BBA {} Schmidt, D\BPBI F.%
\end{APACrefauthors}%
\unskip\
\newblock
\APACrefYearMonthDay{2016}{}{}.
\newblock
{\BBOQ}\APACrefatitle {A Simple Sampler for the Horseshoe Estimator} {A simple sampler for the horseshoe estimator}.{\BBCQ}
\newblock
\APACjournalVolNumPages{IEEE Signal Processing Letters}{23}{1}{179-182}.
\newblock
\begin{APACrefDOI} \doi{10.1109/LSP.2015.2503725} \end{APACrefDOI}
\PrintBackRefs{\CurrentBib}

\bibitem [\protect \citeauthoryear {%
Martin%
\ \BBA {} McDonald%
}{%
Martin%
\ \BBA {} McDonald%
}{%
{\protect \APACyear {1975}}%
}]{%
martin1975}
\APACinsertmetastar {%
martin1975}%
\begin{APACrefauthors}%
Martin, J\BPBI K.%
\BCBT {}\ \BBA {} McDonald, R\BPBI P.%
\end{APACrefauthors}%
\unskip\
\newblock
\APACrefYearMonthDay{1975}{}{}.
\newblock
{\BBOQ}\APACrefatitle {{Bayesian Estimation in Unrestricted Factor Analysis: A Treatment for Heywood Cases}} {{Bayesian Estimation in Unrestricted Factor Analysis: A Treatment for Heywood Cases}}.{\BBCQ}
\newblock
\APACjournalVolNumPages{Psychometrika}{40}{4}{505--517}.
\PrintBackRefs{\CurrentBib}

\bibitem [\protect \citeauthoryear {%
Matuk%
, Herring%
\BCBL {}\ \BBA {} Dunson%
}{%
Matuk%
\ \protect \BOthers {.}}{%
{\protect \APACyear {2025}}%
}]{%
matuk2025}
\APACinsertmetastar {%
matuk2025}%
\begin{APACrefauthors}%
Matuk, J.%
, Herring, A\BPBI H.%
\BCBL {}\ \BBA {} Dunson, D\BPBI B.%
\end{APACrefauthors}%
\unskip\
\newblock
\APACrefYearMonthDay{2025}{}{}.
\newblock
{\BBOQ}\APACrefatitle {Bayesian Modeling of Nearly Mutually Orthogonal Processes} {Bayesian modeling of nearly mutually orthogonal processes}.{\BBCQ}
\newblock
\APACjournalVolNumPages{Bayesian Analysis}{Advance Publication}{}{1--26}.
\newblock
\begin{APACrefDOI} \doi{10.1214/26-BA1600} \end{APACrefDOI}
\PrintBackRefs{\CurrentBib}

\bibitem [\protect \citeauthoryear {%
McLachlan%
, Peel%
\BCBL {}\ \BBA {} Bean%
}{%
McLachlan%
\ \protect \BOthers {.}}{%
{\protect \APACyear {2003}}%
}]{%
mclachlan2003}
\APACinsertmetastar {%
mclachlan2003}%
\begin{APACrefauthors}%
McLachlan, G\BPBI J.%
, Peel, D.%
\BCBL {}\ \BBA {} Bean, R\BPBI W.%
\end{APACrefauthors}%
\unskip\
\newblock
\APACrefYearMonthDay{2003}{}{}.
\newblock
{\BBOQ}\APACrefatitle {Modelling High-Dimensional Data by Mixtures of Factor Analyzers} {Modelling high-dimensional data by mixtures of factor analyzers}.{\BBCQ}
\newblock
\APACjournalVolNumPages{Computational Statistics \& Data Analysis}{41}{3--4}{379--388}.
\PrintBackRefs{\CurrentBib}

\bibitem [\protect \citeauthoryear {%
Michael%
, Schucany%
\BCBL {}\ \BBA {} Haas%
}{%
Michael%
\ \protect \BOthers {.}}{%
{\protect \APACyear {1976}}%
}]{%
michael1976}
\APACinsertmetastar {%
michael1976}%
\begin{APACrefauthors}%
Michael, J\BPBI R.%
, Schucany, W\BPBI R.%
\BCBL {}\ \BBA {} Haas, R\BPBI W.%
\end{APACrefauthors}%
\unskip\
\newblock
\APACrefYearMonthDay{1976}{}{}.
\newblock
{\BBOQ}\APACrefatitle {Generating Random Variates Using Transformations with Multiple Roots} {Generating random variates using transformations with multiple roots}.{\BBCQ}
\newblock
\APACjournalVolNumPages{The American Statistician}{30}{2}{88--90}.
\newblock
\begin{APACrefDOI} \doi{10.2307/2683801} \end{APACrefDOI}
\PrintBackRefs{\CurrentBib}

\bibitem [\protect \citeauthoryear {%
Murray%
, Dunson%
, Carin%
\BCBL {}\ \BBA {} Lucas%
}{%
Murray%
\ \protect \BOthers {.}}{%
{\protect \APACyear {2013}}%
}]{%
murray2013}
\APACinsertmetastar {%
murray2013}%
\begin{APACrefauthors}%
Murray, J\BPBI S.%
, Dunson, D\BPBI B.%
, Carin, L.%
\BCBL {}\ \BBA {} Lucas, J\BPBI E.%
\end{APACrefauthors}%
\unskip\
\newblock
\APACrefYearMonthDay{2013}{}{}.
\newblock
{\BBOQ}\APACrefatitle {Bayesian {G}aussian copula factor models for mixed data} {Bayesian {G}aussian copula factor models for mixed data}.{\BBCQ}
\newblock
\APACjournalVolNumPages{Journal of the American Statistical Association}{108}{502}{656--665}.
\newblock
\begin{APACrefDOI} \doi{10.1080/01621459.2012.762328} \end{APACrefDOI}
\PrintBackRefs{\CurrentBib}

\bibitem [\protect \citeauthoryear {%
Onorati%
, Dunson%
\BCBL {}\ \BBA {} Canale%
}{%
Onorati%
\ \protect \BOthers {.}}{%
{\protect \APACyear {2025}}%
}]{%
onorati2025}
\APACinsertmetastar {%
onorati2025}%
\begin{APACrefauthors}%
Onorati, P.%
, Dunson, D\BPBI B.%
\BCBL {}\ \BBA {} Canale, A.%
\end{APACrefauthors}%
\unskip\
\newblock
\APACrefYearMonthDay{2025}{}{}.
\newblock
{\BBOQ}\APACrefatitle {On the Posterior Computation Under the {Dirichlet-Laplace} Prior} {On the posterior computation under the {Dirichlet-Laplace} prior}.{\BBCQ}
\newblock
\APACjournalVolNumPages{arXiv preprint arXiv:2507.05214}{}{}{}.
\PrintBackRefs{\CurrentBib}

\bibitem [\protect \citeauthoryear {%
Poworoznek%
, Anceschi%
, Ferrari%
\BCBL {}\ \BBA {} Dunson%
}{%
Poworoznek%
\ \protect \BOthers {.}}{%
{\protect \APACyear {2025}}%
}]{%
poworoznek2025}
\APACinsertmetastar {%
poworoznek2025}%
\begin{APACrefauthors}%
Poworoznek, E.%
, Anceschi, N.%
, Ferrari, F.%
\BCBL {}\ \BBA {} Dunson, D.%
\end{APACrefauthors}%
\unskip\
\newblock
\APACrefYearMonthDay{2025}{}{}.
\newblock
{\BBOQ}\APACrefatitle {Efficiently Resolving Rotational Ambiguity in {Bayesian} Matrix Sampling with Matching} {Efficiently resolving rotational ambiguity in {Bayesian} matrix sampling with matching}.{\BBCQ}
\newblock
\APACjournalVolNumPages{Bayesian Analysis}{Advance Publication}{}{1--22}.
\newblock
\begin{APACrefDOI} \doi{10.1214/25-BA1544} \end{APACrefDOI}
\PrintBackRefs{\CurrentBib}

\bibitem [\protect \citeauthoryear {%
Press%
\ \BBA {} Shigemasu%
}{%
Press%
\ \BBA {} Shigemasu%
}{%
{\protect \APACyear {1989}}%
}]{%
press1989}
\APACinsertmetastar {%
press1989}%
\begin{APACrefauthors}%
Press, S\BPBI J.%
\BCBT {}\ \BBA {} Shigemasu, K.%
\end{APACrefauthors}%
\unskip\
\newblock
\APACrefYearMonthDay{1989}{}{}.
\newblock
{\BBOQ}\APACrefatitle {{Bayesian Inference in Factor Analysis}} {{Bayesian Inference in Factor Analysis}}.{\BBCQ}
\newblock
\BIn{} L.~Gleser, M.~Perlman, S\BPBI J.~Press\BCBL {}\ \BBA {} A.~Sampson\ (\BEDS), \APACrefbtitle {Contributions to Probability and Statistics: Essays in Honor of Ingram Olkin} {Contributions to probability and statistics: Essays in honor of ingram olkin}\ (\BPGS\ 271--287).
\newblock
\APACaddressPublisher{}{Springer}.
\PrintBackRefs{\CurrentBib}

\bibitem [\protect \citeauthoryear {%
{R Core Team}%
}{%
{R Core Team}%
}{%
{\protect \APACyear {2025}}%
}]{%
Rmanual}
\APACinsertmetastar {%
Rmanual}%
\begin{APACrefauthors}%
{R Core Team}.%
\end{APACrefauthors}%
\unskip\
\newblock
\APACrefYearMonthDay{2025}{}{}.
\newblock
{\BBOQ}\APACrefatitle {R: A Language and Environment for Statistical Computing} {R: A language and environment for statistical computing}{\BBCQ}\ [\bibcomputersoftwaremanual].
\newblock
\APACaddressPublisher{Vienna, Austria}{}.
\newblock
\begin{APACrefURL} \url{https://www.R-project.org/} \end{APACrefURL}
\PrintBackRefs{\CurrentBib}

\bibitem [\protect \citeauthoryear {%
Rockova%
\ \BBA {} George%
}{%
Rockova%
\ \BBA {} George%
}{%
{\protect \APACyear {2018}}%
}]{%
rockova2018}
\APACinsertmetastar {%
rockova2018}%
\begin{APACrefauthors}%
Rockova, V.%
\BCBT {}\ \BBA {} George, E\BPBI I.%
\end{APACrefauthors}%
\unskip\
\newblock
\APACrefYearMonthDay{2018}{}{}.
\newblock
{\BBOQ}\APACrefatitle {{The Spike-and-Slab LASSO}} {{The Spike-and-Slab LASSO}}.{\BBCQ}
\newblock
\APACjournalVolNumPages{Journal of the American Statistical Association}{113}{521}{431--444}.
\PrintBackRefs{\CurrentBib}

\bibitem [\protect \citeauthoryear {%
Rubin%
\ \BBA {} Thayer%
}{%
Rubin%
\ \BBA {} Thayer%
}{%
{\protect \APACyear {1982}}%
}]{%
rubin1982}
\APACinsertmetastar {%
rubin1982}%
\begin{APACrefauthors}%
Rubin, D\BPBI B.%
\BCBT {}\ \BBA {} Thayer, D\BPBI T.%
\end{APACrefauthors}%
\unskip\
\newblock
\APACrefYearMonthDay{1982}{}{}.
\newblock
{\BBOQ}\APACrefatitle {{EM Algorithms for ML Factor Analysis}} {{EM Algorithms for ML Factor Analysis}}.{\BBCQ}
\newblock
\APACjournalVolNumPages{Psychometrika}{47}{1}{69--76}.
\PrintBackRefs{\CurrentBib}

\bibitem [\protect \citeauthoryear {%
Schiavon%
, Canale%
\BCBL {}\ \BBA {} Dunson%
}{%
Schiavon%
\ \protect \BOthers {.}}{%
{\protect \APACyear {2022}}%
}]{%
schiavon2022}
\APACinsertmetastar {%
schiavon2022}%
\begin{APACrefauthors}%
Schiavon, L.%
, Canale, A.%
\BCBL {}\ \BBA {} Dunson, D\BPBI B.%
\end{APACrefauthors}%
\unskip\
\newblock
\APACrefYearMonthDay{2022}{}{}.
\newblock
{\BBOQ}\APACrefatitle {Generalized Infinite Factorization Models} {Generalized infinite factorization models}.{\BBCQ}
\newblock
\APACjournalVolNumPages{Biometrika}{109}{3}{817--835}.
\newblock
\begin{APACrefDOI} \doi{10.1093/biomet/asab056} \end{APACrefDOI}
\PrintBackRefs{\CurrentBib}

\bibitem [\protect \citeauthoryear {%
Schmidt%
\ \BBA {} Gelfand%
}{%
Schmidt%
\ \BBA {} Gelfand%
}{%
{\protect \APACyear {2003}}%
}]{%
schmidt2003}
\APACinsertmetastar {%
schmidt2003}%
\begin{APACrefauthors}%
Schmidt, A\BPBI M.%
\BCBT {}\ \BBA {} Gelfand, A\BPBI E.%
\end{APACrefauthors}%
\unskip\
\newblock
\APACrefYearMonthDay{2003}{}{}.
\newblock
{\BBOQ}\APACrefatitle {A {Bayesian} Coregionalization Approach for Multivariate Pollutant Data} {A {Bayesian} coregionalization approach for multivariate pollutant data}.{\BBCQ}
\newblock
\APACjournalVolNumPages{Journal of Geophysical Research: Atmospheres}{108}{D24}{8783}.
\newblock
\begin{APACrefDOI} \doi{10.1029/2002JD002905} \end{APACrefDOI}
\PrintBackRefs{\CurrentBib}

\bibitem [\protect \citeauthoryear {%
Spearman%
}{%
Spearman%
}{%
{\protect \APACyear {1904}}%
}]{%
spearman1904}
\APACinsertmetastar {%
spearman1904}%
\begin{APACrefauthors}%
Spearman, C.%
\end{APACrefauthors}%
\unskip\
\newblock
\APACrefYearMonthDay{1904}{}{}.
\newblock
{\BBOQ}\APACrefatitle {{General Intelligence, Objectively Determined and Measured}} {{General Intelligence, Objectively Determined and Measured}}.{\BBCQ}
\newblock
\APACjournalVolNumPages{The American Journal of Psychology}{15}{2}{201--292}.
\PrintBackRefs{\CurrentBib}

\bibitem [\protect \citeauthoryear {%
Yalcin%
\ \BBA {} Amemiya%
}{%
Yalcin%
\ \BBA {} Amemiya%
}{%
{\protect \APACyear {2001}}%
}]{%
yalcin2001}
\APACinsertmetastar {%
yalcin2001}%
\begin{APACrefauthors}%
Yalcin, I.%
\BCBT {}\ \BBA {} Amemiya, Y.%
\end{APACrefauthors}%
\unskip\
\newblock
\APACrefYearMonthDay{2001}{}{}.
\newblock
{\BBOQ}\APACrefatitle {Nonlinear factor analysis as a statistical method} {Nonlinear factor analysis as a statistical method}.{\BBCQ}
\newblock
\APACjournalVolNumPages{Statistical Science}{}{}{275--294}.
\PrintBackRefs{\CurrentBib}

\bibitem [\protect \citeauthoryear {%
Zhao%
, Gao%
, Mukherjee%
\BCBL {}\ \BBA {} Engelhardt%
}{%
Zhao%
\ \protect \BOthers {.}}{%
{\protect \APACyear {2016}}%
}]{%
zhao2016}
\APACinsertmetastar {%
zhao2016}%
\begin{APACrefauthors}%
Zhao, S.%
, Gao, C.%
, Mukherjee, S.%
\BCBL {}\ \BBA {} Engelhardt, B\BPBI E.%
\end{APACrefauthors}%
\unskip\
\newblock
\APACrefYearMonthDay{2016}{}{}.
\newblock
{\BBOQ}\APACrefatitle {{Bayesian Group Factor Analysis with Structured Sparsity}} {{Bayesian Group Factor Analysis with Structured Sparsity}}.{\BBCQ}
\newblock
\APACjournalVolNumPages{Journal of Machine Learning Research}{17}{196}{1--47}.
\PrintBackRefs{\CurrentBib}

\end{thebibliography}

\end{singlespace}

\newpage

\section{Supplementary}

\subsection{Supplementary C++ Implementation}\label{subsec:sup_c++}

\subsubsection{Dirichlet--Laplace Prior Updates}\label{subsubsec:DL_post}

The DL prior of \cite{bhattacharya2015} replaces the MGPS shrinkage parameters with a Dirichlet--Laplace structure on the loadings. The key difference is that instead of the cumulative shrinkage $\tau_k$, we use local scales $\psi_{pk}$, Dirichlet-distributed weights $\phi_{pk}$, and row-specific scales $\tau_p$. The factor score and residual precision updates remain identical to Section~\ref{subsubsec:likelihood}. The three shrinkage blocks are sampled in the order $\phi$, then $\tau$, then $\psi$, so that each draw conditions on the freshly updated values of the preceding blocks; sampling in a different order does not target the correct full conditional, as recently clarified by \citet{onorati2025} and in the published correction of \citet{gruber2025}.

{\it Sampling Dirichlet Weights.}
The weights $\phi_{pk}$ are sampled row-wise from a generalized Dirichlet distribution. For each row $p$, we sample auxiliary variables $T_k$ from GIG distributions and normalize:

\begin{minted}[mathescape, baselinestretch=1.2, bgcolor=LightGray, fontsize=\footnotesize]{cpp}
for (int p = 0; p < P; ++p) {
    arma::vec T(K);
    // Sample auxiliary variables from GIG
    for (int k = 0; k < K; ++k)
        T(k) = rgig(a - 1.0, 2.0 * std::abs(lambda(p,k)), 1.0);
    
    // Normalize to obtain Dirichlet weights
    double Tsum = arma::sum(T);
    if (Tsum > 0) {
        phi.row(p) = (T / Tsum).t();
    } else {
        phi.row(p).fill(1.0/K);
    }
}
\end{minted}

{\it Sampling Row-Specific Scales.}
Each row scale $\tau_p$ is sampled from a generalized inverse Gaussian (GIG) distribution with parameters depending on the sum of $|\lambda_{pk}|/\phi_{pk}$ across that row:

\begin{minted}[mathescape, baselinestretch=1.2, bgcolor=LightGray, fontsize=\footnotesize]{cpp}
// For each row p, sample tau_p from GIG(a*K - K, 2*sum_k |lambda_pk|/phi_pk, 1)
for (int p=0; p<P; ++p) {
    double num = 0.0;
    for (int k=0; k<K; ++k)
        num += std::abs(lambda(p,k)) / phi(p,k);
    tau(p) = rgig(a*K - K, 2.0*num, 1.0);
}
\end{minted}

{\it Sampling Local Scales.}
The local scales $\psi_{pk}$ are sampled from their inverse Gaussian full conditional with mean $\mu_{pk} = \phi_{pk}\tau_p/|\lambda_{pk}|$ and shape $\lambda=1$. The \texttt{C++} code uses the \cite{michael1976} algorithm:

\begin{minted}[mathescape, baselinestretch=1.2, bgcolor=LightGray, fontsize=\footnotesize]{cpp}
// Loop through all loadings and sample local scales:
for (int p=0; p<P; ++p) {
    for (int k=0; k<K; ++k) {
        // Mean of inverse Gaussian: phi_pk * tau_p / |lambda_pk|
        double mu = phi(p,k) * tau(p) / (std::abs(lambda(p,k)) + 1e-10);
        
        // Sample from inverse Gaussian and invert
        psi(p,k) = 1.0 / rinvgauss(mu, 1.0);
    }
}
\end{minted}

{\it Constructing the Prior Precision.}
The Dirichlet--Laplace components define the prior \emph{variance} of each loading as $\sigma^2_{\lambda,pk} = \psi_{pk}\,\phi_{pk}^2\,\tau_p^2$, following the scale-mixture representation of \cite{bhattacharya2015}. The loadings update in Section~\ref{subsubsec:likelihood} works with the prior precision, so we pass the reciprocal:

\begin{minted}[mathescape, baselinestretch=1.2, bgcolor=LightGray, fontsize=\footnotesize]{cpp}
// Prior variance: psi_pk * (phi_pk * tau_p)^2 ; precision is its reciprocal
arma::mat phitau  = phi.each_col() % tau;      // phi_pk * tau_p (per row)
arma::mat var_lam = psi % (phitau % phitau);   // psi * (phi*tau_p)^2
Plam = 1.0 / var_lam;                          // precision = 1 / variance
\end{minted}

The matrix $\texttt{Plam}$ holds the prior precisions $1/(\psi_{pk}\phi_{pk}^2\tau_p^2)$, replacing the MGPS precision $\mathrm{diag}(\boldsymbol{\phi}_{p\cdot}\boldsymbol{\tau})$ in the loadings update from Section~\ref{subsubsec:likelihood}, while all other updates (factor scores, residual precisions) proceed identically.

\subsubsection{Horseshoe Prior Updates}\label{subsubsec:HS_post}

The Horseshoe prior replaces the MGPS shrinkage parameters with a half-Cauchy structure using the \cite{makalic2016} auxiliary variable representation. The key parameters are local scales $\phi^2_{pk}$, a global scale $\tau^2$, and auxiliary variables $\nu_{pk}$ and $\xi$ that enable Gibbs sampling. The factor score and residual precision updates remain identical to Section~\ref{subsubsec:likelihood}. 

{\it Sampling Local Scales.}
The local scales $\tau^2_{pk,\text{local}}$ are sampled from inverse Gamma full conditionals with shape $1$ and rate depending on the loading magnitude and auxiliary variable $\xi_{pk}$:

\begin{minted}[mathescape, baselinestretch=1.2, bgcolor=LightGray, fontsize=\footnotesize]{cpp}
// Loop through all loadings and sample local scales:
for(int p=0; p<P; ++p) {
    for(int k=0; k<K; ++k) {
        double shape = 1.0;
        double rate  = std::pow(lambda(p,k),2)/(2.0*tau2_global) + 1.0/xi_local(p,k);
        tau2_local(p,k) = 1.0 / R::rgamma(shape, 1.0 / rate);
    }
}
\end{minted}

{\it Sampling Local Auxiliary Variables.}
The auxiliary variables $\xi_{pk}$ are sampled from inverse Gamma distributions that induce the half-Cauchy prior on $\tau_{pk,\text{local}}$:

\begin{minted}[mathescape, baselinestretch=1.2, bgcolor=LightGray, fontsize=\footnotesize]{cpp}
// Loop through all entries and sample auxiliary variables:
for(int p=0; p<P; ++p)
    for(int k=0; k<K; ++k)
        xi_local(p,k) = 1.0 / R::rgamma(1.0, 1.0 + 1.0 / tau2_local(p,k));
\end{minted}

{\it Sampling Global Scale.}
The global scale $\tau^2_{\text{global}}$ is sampled from an inverse Gamma full conditional with shape $(PK+1)/2$ and rate depending on the sum of squared loadings scaled by local variances:

\begin{minted}[mathescape, baselinestretch=1.2, bgcolor=LightGray, fontsize=\footnotesize]{cpp}
// Compute sum of lambda_pk^2 / tau2_local_pk
double sum_term = accu(square(lambda) / tau2_local);

// Sample from inverse Gamma
double shape = 0.5 * (P*K + 1.0);
double rate  = 0.5 * sum_term + 1.0 / xi_global;
tau2_global = 1.0 / R::rgamma(shape, 1.0 / rate);
\end{minted}

{\it Sampling Global Auxiliary Variable.}
The auxiliary variable $\xi_{\text{global}}$ is sampled similarly to induce the half-Cauchy prior on $\tau_{\text{global}}$:

\begin{minted}[mathescape, baselinestretch=1.2, bgcolor=LightGray, fontsize=\footnotesize]{cpp}
xi_global = 1.0 / R::rgamma(1.0, 1.0 + 1.0 / tau2_global);
\end{minted}

{\it Constructing Prior Precision.}
The prior precision matrix for the loadings is constructed as $\texttt{Plam}_{pk} = 1/(\tau^2_{\text{global}} \cdot \tau^2_{pk,\text{local}})$:

\begin{minted}[mathescape, baselinestretch=1.2, bgcolor=LightGray, fontsize=\footnotesize]{cpp}
// Build the precision matrix: Plam_pk = 1 / (tau2_global * tau2_local_pk)
Plam = 1.0 / (tau2_global * tau2_local);
\end{minted}

The matrix $\texttt{Plam}$ replaces the MGPS precision $\boldsymbol{\phi}_{p\cdot}\boldsymbol{\tau}$ in the loadings update from Section~\ref{subsubsec:likelihood}, while all other updates (factor scores, residual precisions) proceed identically.

\subsubsection{BASS Prior Updates}\label{subsubsec:BASS_post}

The BASS prior of \citet{zhao2016} replaces the MGPS shrinkage parameters with a three-level structured shrinkage hierarchy combined with a per-factor sparse/dense mixture. Each loading $\lambda_{pk}$ is given a local precision $\theta_{pk}$, which is in turn governed by a chain of higher-level scales $\delta_{pk}$, $\omega_k$, $\rho_k$, $\eta$, and $\gamma$, each pair forming a three-parameter beta (horseshoe-type) relationship. A factor-level indicator $z_k$ selects, for each factor, between a \emph{sparse} regime, in which the local precisions are learned element by element, and a \emph{dense} regime, in which all loadings in that factor share a common scale. The factor score and residual precision updates remain identical to Section~\ref{subsubsec:likelihood}. 

{\it Sampling Factor Indicators.}
For each factor $k$, the indicator $z_k$ is drawn from its Bernoulli full conditional, comparing the marginal evidence for the sparse element-wise model against the dense common-scale model, weighted by the mixture proportion $\pi$.

The \texttt{C++} code is

\begin{minted}[mathescape, baselinestretch=1.2, bgcolor=LightGray, fontsize=\footnotesize]{cpp}
for(int k=0; k<K; ++k) {
    // Log evidence for sparse (z=1) vs dense (z=0), weighted by pi
    double log_sparse = std::log(pi);
    double log_dense  = std::log(1.0 - pi);
    for(int p=0; p<P; ++p) {
        double lam2 = std::pow(lambda(p,k), 2);
        log_sparse += (a + 0.5) * std::log(delta(p,k) + 0.5*lam2)
                    - std::log(delta(p,k) + 0.5*lam2) * (1.0/(delta(p,k) + 0.5*lam2));
        log_dense  += -0.5 * std::log(phi(k)) - 0.5 * lam2 * phi(k);
    }
    double prob_sparse = 1.0 / (1.0 + std::exp(log_dense - log_sparse));
    out(k) = (R::runif(0,1) < prob_sparse) ? 1 : 0;
}
\end{minted}

{\it Sampling the Mixture Proportion.}
The proportion of sparse factors $\pi$ is drawn from its Beta full conditional under a $\text{Beta}(1,1)$ prior, where $\sum_k z_k$ counts the sparse factors:

\begin{minted}[mathescape, baselinestretch=1.2, bgcolor=LightGray, fontsize=\footnotesize]{cpp}
int n_sparse = arma::accu(z);
pi = R::rbeta(1.0 + n_sparse, 1.0 + K - n_sparse);
\end{minted}

{\it Sampling Local Precisions.}
The local precisions $\theta_{pk}$ are drawn from Gamma full conditionals. For a sparse factor ($z_k=1$) each $\theta_{pk}$ is updated element-wise with rate $\delta_{pk} + \tfrac{1}{2}\lambda_{pk}^2$; for a dense factor ($z_k=0$) every loading in the factor takes the common value $\omega_k$:

\begin{minted}[mathescape, baselinestretch=1.2, bgcolor=LightGray, fontsize=\footnotesize]{cpp}
for(int k=0; k<K; ++k) {
    if (z(k) == 1) {                       // sparse: element-wise
        for(int p=0; p<P; ++p)
            theta(p,k) = R::rgamma(a + 0.5,
                          1.0/(delta(p,k) + 0.5*std::pow(lambda(p,k),2)));
    } else {                               // dense: shared scale
        theta.col(k).fill(phi(k));
    }
}
\end{minted}

{\it Sampling Higher-Level Scales.}
The remaining levels of the hierarchy are conjugate Gamma updates, each scale governed by the level above it. The variable $\delta_{pk}$ is governed by $\omega_k$, $\omega_k$ by $\rho_k$, $\rho_k$ by $\eta$, and $\eta$ by $\gamma$:

\begin{minted}[mathescape, baselinestretch=1.2, bgcolor=LightGray, fontsize=\footnotesize]{cpp}
// delta_{pk} | omega_k
for(int p=0;p<P;++p)
    for(int k=0;k<K;++k)
        delta(p,k) = R::rgamma(b + 1.0, 1.0/(omega(k) + theta(p,k)));

// omega_k | rho_k
for(int k=0;k<K;++k)
    omega(k) = R::rgamma(c + P, 1.0/(rho(k) + arma::accu(delta.col(k))));

// rho_k | eta
for(int k=0;k<K;++k)
    rho(k) = R::rgamma(d + 1.0, 1.0/(eta + omega(k)));

// eta | gamma  and  gamma | nu
eta   = R::rgamma(e + K, 1.0/(gamma + arma::accu(rho)));
gamma = R::rgamma(f + 1.0, 1.0/(nu + eta));
\end{minted}

{\it Sampling Factor Loadings.} 
The loadings update is identical to Section~\ref{subsubsec:likelihood}, with the prior precision matrix taken to be $\mathrm{diag}(\theta_{p1},\ldots,\theta_{pK})$, so that $\boldsymbol{M}_p = \sigma_p^{-2}\sum_i \mathbf{b}_i\mathbf{b}_i^t + \mathrm{diag}(\boldsymbol{\theta}_{p\cdot})$. All hyperparameters are set to $1/2$ (a horseshoe at every level), with $\nu=1$. As with the other priors, the factor score and residual precision updates proceed exactly as in Section~\ref{subsubsec:likelihood}.

\subsubsection{MNL Score Prior Updates}\label{subsubsec:MNL_post}

The MNL prior of \citet{huang2025} differs more substantially from the others: rather than placing a shrinkage prior on the loadings, it places a mass-nonlocal (spike-and-pMOM) prior on the factor \emph{scores} $\boldsymbol{b}$, and retains the MGPS prior on the loadings. Each score $b_{ik}$ is drawn from a spike at zero with probability $1-\theta_k$, or from a product-moment (pMOM) slab with probability $\theta_k$, where $\theta_k$ controls the proportion of nonzero scores on factor $k$. The loading, local shrinkage, and global multiplier updates are exactly those of the MGPS prior (Section~\ref{subsubsec:mgps_post}); the residual precision update is as in Section~\ref{subsubsec:likelihood}. We describe only the components specific to MNL: the latent indicators, the scores, and the score-sparsity parameters.

Our implementation is a faithful port of the authors' reference code; on data generated from the model it matches the reference RV coefficient to within Monte Carlo error.

{\it Sampling Latent Indicators.}
The indicator $Z_{ik}$ records whether score $b_{ik}$ is drawn from the slab ($Z_{ik}=1$) or the spike ($Z_{ik}=0$). Its full conditional marginalizes the score $b_{ik}$ in closed form. With $a = 1/\psi_k + \boldsymbol{\lambda}_k^t\boldsymbol{\Psi}\boldsymbol{\lambda}_k$ and $d = (\mathbf{x}_i - \boldsymbol{\Lambda}_{-k}\mathbf{b}_{i,-k})^t \boldsymbol{\Psi}\boldsymbol{\lambda}_k$, the slab-to-spike evidence ratio is $t = (a\psi_k)^{-3/2}(1 + d^2/a)\exp\{d^2/(2a)\}$ and $\Pr(Z_{ik}=1) = 1 - (1-\theta_k)/(1-\theta_k+\theta_k t)$. The \texttt{C++} code is

\begin{minted}[mathescape, baselinestretch=1.2, bgcolor=LightGray, fontsize=\footnotesize]{cpp}
for (int i = 0; i < n; ++i) {
    for (int k = 0; k < K; ++k) {
        // a = 1/psi_k + lambda_k' Psi lambda_k
        double a = 1.0 / psi(k);
        for (int j = 0; j < P; ++j)
            a += Sigma_inv(j) * Lambda(j,k) * Lambda(j,k);
        // d = (x_i - Lambda_{-k} b_{i,-k})' Psi lambda_k
        double d = 0.0;
        for (int j = 0; j < P; ++j) {
            double r = X(i,j);
            for (int l = 0; l < K; ++l)
                if (l != k) r -= Lambda(j,l) * b(i,l);
            d += r * Sigma_inv(j) * Lambda(j,k);
        }
        double t    = std::pow(a*psi(k), -1.5) * (1.0 + d*d/a) * std::exp(d*d/(2.0*a));
        double prob = 1.0 - (1.0 - theta(k)) / (1.0 - theta(k) + theta(k)*t);
        if (prob < 0.0) prob = 0.0;
        if (prob > 1.0) prob = 1.0;
        Z(i,k) = R::rbinom(1, prob);
    }
}
\end{minted}

{\it Sampling Factor Scores.}
When $Z_{ik}=0$ the score is set to zero. When $Z_{ik}=1$ the score is drawn from its pMOM full conditional, proportional to $\exp\{-a(b + d/a)^2\}\,b^2$, with $a = \tfrac{1}{2}(\boldsymbol{\lambda}_k^t\boldsymbol{\Psi}\boldsymbol{\lambda}_k + 1/\psi_k)$ and $d = \tfrac{1}{2}\boldsymbol{\lambda}_k^t\boldsymbol{\Psi} (\boldsymbol{\Lambda}_{-k}\mathbf{b}_{i,-k} - \mathbf{x}_i)$. Because this density is not of standard form, it is sampled with a short random-walk Metropolis--Hastings step (two iterations, unit step size), matching the reference implementation:

\begin{minted}[mathescape, baselinestretch=1.2, bgcolor=LightGray, fontsize=\footnotesize]{cpp}
// log target: -a*(x + d/a)^2 + log(x^2)
inline double log_pos_b(double x, double a, double d) {
    return -a * std::pow(x + d/a, 2) + std::log(x*x + 1e-300);
}
inline double mh_b(double init, double a, double d, int M = 2, double step = 1.0) {
    double cur = init;
    for (int m = 1; m < M; ++m) {
        double cand = R::rnorm(cur, step);
        double logr = log_pos_b(cand, a, d) - log_pos_b(cur, a, d);
        if (std::log(R::runif(0.0, 1.0)) <= logr) cur = cand;
    }
    return cur;
}
\end{minted}

The score update applies \texttt{mh\_b} to each $b_{ik}$ with $Z_{ik}=1$, using the $a$ and $d$ above, and sets $b_{ik}=0$ otherwise.

{\it Sampling Score-Sparsity Parameters.}
The proportion of nonzero scores on each factor, $\theta_k$, is drawn from its Beta full conditional under a $\text{Beta}(1,5)$ prior, counting the slab allocations in column $k$ of $\mathbf{Z}$. The pMOM dispersion $\psi_k$ is then set deterministically as $\psi_k = 1/(3\theta_k)$, as in the reference:

\begin{minted}[mathescape, baselinestretch=1.2, bgcolor=LightGray, fontsize=\footnotesize]{cpp}
for (int k = 0; k < K; ++k) {
    double sumZ = arma::accu(Z.col(k));
    theta(k) = R::rbeta(1.0 + sumZ, 5.0 + (double)n - sumZ);
}
psi = 1.0 / (3.0 * theta);
\end{minted}

The loadings, local shrinkage $\phi_{pk}$, global multipliers $\delta_k$, and residual precisions are updated exactly as for the MGPS prior. Posterior summaries for $\boldsymbol{\Lambda}$ and $\boldsymbol{b}$ use the median over draws, following the reference implementation.

\end{document}